\documentclass[10pt,journal,compsoc]{IEEEtran}
\IEEEoverridecommandlockouts
\usepackage{optidef}
\usepackage{cite}
\usepackage{amsmath,amssymb,amsfonts}
\usepackage{algorithmic}
\usepackage{graphicx}
\usepackage{textcomp}
\usepackage{xcolor}
\usepackage{stfloats}
\usepackage[nodisplayskipstretch]{setspace}
\def\BibTeX{{\rm B\kern-.05em{\sc i\kern-.025em b}\kern-.08em
    T\kern-.1667em\lower.7ex\hbox{E}\kern-.125emX}}
  
\usepackage{bbm}
\usepackage{caption,setspace}
\usepackage[font=small]{caption}
\usepackage[font=footnotesize]{caption}
\newtheorem{remark}{Remark}
\begin{document}
\setlength{\belowdisplayskip}{0pt} \setlength{\belowdisplayshortskip}{0pt}
\setlength{\abovedisplayskip}{0pt} \setlength{\abovedisplayshortskip}{0pt}

\title{Content-Aware User Association and Multi-User MIMO Beamforming over Mobile Edge Caching}


\author{Susanna~Mosleh,
        Qiang~Fan,
        Jonathan~D.~Ashdown,~\IEEEmembership{Member,~IEEE,}
        Erik~Perrins,~\IEEEmembership{Senior~Member,~IEEE,}
        Kurt~Turck,
        and~Lingjia~Liu,~\IEEEmembership{Senior~Member,~IEEE}
\IEEEcompsocitemizethanks{\IEEEcompsocthanksitem S. Mosleh is with the Department of Physics, University of Colorado, Boulder, and an associate of the Communications Technology Laboratory, National Institute of Standards and Technology, Boulder, CO, 80305.\protect\\
E-mail: susanna.mosleh@colorado.edu
\IEEEcompsocthanksitem Q. Fan and L. Liu are with the Department of Electrical and Computer Engineering, Virginia Tech, Blacksburg, VA, 24061.\protect\\
E-mail: \{qiangf,~ljliu\}@vt.edu
\IEEEcompsocthanksitem E. Perrins is with the Department of Electrical Engineering and Computer Science, the University of Kansas, Lawrence, KS, 66045. \protect\\
E-mail: esp@ku.edu
\IEEEcompsocthanksitem J. Ashdown and K. Turck are with the Information Directorate, Air Force Research Laboratory (AFRL), Rome, NY, 13441.. \protect\\
E-mail:\{jonathan.ashdown,~kurt.turck\}@us.af.mil}
\thanks{The corresponding author is L. Liu (ljliu@ieee.org).}}

\IEEEtitleabstractindextext{
\begin{abstract}
Mobile edge caching (MEC) has been introduced to support ever-growing end-users' needs. To reduce the backhaul traffic demand and content delivery latency, cache-enabled edge servers at base stations (BSs) are employed to provision popular contents at the network edge. In this paper, multiple-input-multiple-output (MIMO) operation and user association policy are linked to the underlying cache placement strategy to ensure a good trade-off between load balancing and backhaul traffic taking into account the underlying wireless channel and the finite cache capacity at edge servers. Due to the coupled interference among mobile stations, the binary nature of the underlying cache placement and user association matrices, the resulting mixed-timescale mixed integer optimization problem is nonconvex and NP-hard. To solve this problem, we decompose the joint optimization problem into a long-term content placement sub-problem and a short-term content delivery sub-problem. A novel iterative algorithm is introduced by leveraging the alternating direction method of multipliers together with a stochastic parallel successive convex approximation-based algorithm. The introduced scheme enables all BSs to update their optimization variables in parallel by solving a sequence of convex subproblems. Simulation evaluation demonstrates the efficiency of our strategy. 
\end{abstract}

}

\maketitle

\section{Introduction}
\IEEEPARstart{T}{he} big data era is being shaped with the ongoing growth
of commercial data services, with mobile wireless network constituting a major data source contributor. Nowadays,
wireless communication is becoming tightly integrated in our
daily lives; especially with the global spread of laptops, tablets,
smartphones, video streaming and online social networking
applications. This globalization has paved the way to dramatically increase wireless network dimensions in terms of
subscribers and the amount of flowing data.  Cisco Systems forecasts that the number of mobile-connected devices per person will reach 1.5 by 2022 and
global mobile data traffic will increase seven fold between
2017 and 2022 \cite{Cisco_VNI2017}. The volume, velocity, and variety of data
from both mobile users and communication networks follows an exponential increase pattern. Consequently, big data will
be further entrenched in the upcoming fifth-generation (5G) wireless networks in which the ability to support high data traffic and exceedingly low latency play an important fundamental role \cite{Mosleh-2017}.

In traditional mobile network, user content requests are served by Internet content providers. Once certain contents are required, they can only be retrieved from the remote cloud or content servers. Since some popular contents may be requested for multiple times, the content provider has to repeat to transfer the identical content. The repeated requests lead to long service latency as well as increasing traffic that may congest the network. To solve this problem, mobile edge computing has been introduced to improve the network performance by provision computing capability at the network edge (i.g. at base stations (BSs)) \cite{fan2019on}. As edge servers can not only provide computing resources but also storage resources, it can be utilized as cache nodes to store valuable contents required by mobile users. Hence, mobile edge caching (MEC) is referred to as caching at edge servers to reduce the traffic at the backhaul and the content delivery latency \cite{Wang2017a}. Note that mobile edge caching has been defined as one of the applications of mobile edge computing by ETSI ISG \cite{etsi2016mobile}. It is promising to store popular contents in cache-enabled edge servers in advance such that users can retrieve contents from the cache in their vicinity rather than the remote cloud. As a result, the cost of the traffic in the backhaul can be remarkably saved, while the content delivery time is improved.

In the MEC with densely deployed BSs, coordinated Multi-Point (CoMP) transmission can be utilized to facilitate the wireless cache \cite{Maddah-2013,Lau-2013, Sengupta-2014, Paschos-2016, Tran-2016, Hajisami-2017}. To do so, a cluster of BSs is assigned to each user to effectively relieve the backhaul capacity demand at MEC. Each cluster is formed by aggregating the BSs whose transmission strategies cooperatively serve the user equipments (UEs) within the cluster through joint proceding \cite{Maddah-2013}. If the user's serving BSs cache the content that the user requests, it will be transmitted directly by the serving BSs cooperatively.  CoMP opportunities also allow a reduction in inter-cell interference \cite{Mosleh-2016}, and thus incur a substantial data rate improvement for users. 

In order to entirely utilize the benefit of MEC with CoMP transmission, developing advanced caching placement strategies in MEC is required. One way to increase the possibility for a user to access its desired content locally is to cache contents based on the data popularity distribution, such as the Zipf distribution \cite{Golrezaei-2014}. Moreover, the content placement has to take user association into account. To be specific, if edge servers have the content requested by a user with undesirable channel gain, data transmission is not reliable and necessitates high transmit power. Consequently, reallocating the requested content via backhaul to BSs that have good channels to the served users will be unavoidable and increase the backhaul cost \cite{Mosleh-2017}. Owing to CoMP technology, each BS has multiple antennas and thus several BSs in a cluster can cooperatively transmit contents to a particular user with multicast beamforming technologies. Therefore, the beamforming vector design also impacts the backhaul cost. In a densely deployed wireless network, each user can be associated with one or several BSs depending on both content availability and channel condition, and thus jointly optimizing the user-association policy, caching placement strategy, and beamforming design remains a challenging issue. 

\subsection{Related Work}

The importance of caching in $5$G wireless networks was recognized in \cite{Leung-2014, Bastug-2014, Boccardi-2014}. Due to the storage resources provided by mobile edge servers, MEC becomes an important use case to reduce the content delivery latency in wireless network\cite{beck2014mobile}. Yao \textit{et al.} have surveyed and investigated the cache placement and cache criteria in MEC to improve both the backhaul cost and content delivery latency in wireless networks \cite{yao2019on}. In \cite{Dai-2014, Zhuang-2014, Peng-2014}, the authors considered the problem of jointly minimizing the total transmit power and backhaul traffic in wireless cooperative networks under the constraint of each user's SINR requirements and with respect to the beamforming vectors. Assuming there is a backhaul constraint for each BS, \cite{Yu-2014} considered a weighted sum rate optimization problem to design the beamforming vectors. In \cite{huang2020online}, an online algorithm on user-AP association and resource allocation for content delivery with predictive scheduling is proposed to improve the throughput, queue stability and delay reduction, where the content placement is fixed. Other than these works on unicast, \cite{Yu-2016,Tao-2016} discussed the effect of caching on the multicast beamforming in the Cloud-RAN. However, these works assumed that the cache placement is static, which means the caching placement matrix is fixed and known at the cloud.

In order to improve the efficiency of cache-enabled networks, \cite{Dimakis-2012,Shanmugam-2013, Li-2015, Mosleh-2017} conducted an investigation into the design of caching policy. In \cite{Rezvani2020fairness}, an fairness-aware cache placement and delivery strategy for an OFDMA-based heterogeneous cellular networks is proposed to improve the network performance, which consists of two separated phases: caching phase and delivery phase. In \cite{Dimakis-2012,Shanmugam-2013}, the caching problem at the small cell BSs was considered, and a caching policy was designed such that the cache-hit-ratio is maximized. To minimize the downloading latency, \cite{Li-2015} proposed a distributed caching algorithm. The line of works in \cite{Dimakis-2012, Shanmugam-2013, Li-2015} are further expanded in \cite{Mosleh-2017} for a cache-enabled Cloud-RAN to consider the trade-off between the transmission power and the backhaul cost of the Cloud-RAN system as a normalized
weighted sum and, minimized the network cost with respect to
both the beamforming matrix and the cache placement matrix
by considering the quality of service (QoS), peak transmission
power, and cache capacity constraints. However, these works are only focused on designing the beamformers and cache placement matrix while assuming the user association matrix is given and known at the cloud.

The user association problem that is mainly concerned with load balancing, was discussed in \cite{Rong-2013, Shen-2014}. The key point here is accounting for both wireless channels and the number of UEs connected to each BS. Assuming caching policy is given, \cite{Pantisano-2014} designed the user association policy such that maximizes the average download rate. These studies, however, did not optimize the user association and cache placement jointly. As a result, the system was led to an inefficient operating point.

Designing jointly the caching strategy and the user association policy in cache-enabled wireless networks is considered in \cite{Poularakis-2014,Iosifidis-2014,Naveen-2015,Dehghan-2015,Khreishah-2016,Wang-2016}. To minimize the number of requests performed by the macro BSs in a small-cell network, \cite{Poularakis-2014} designed a joint data caching and user association policy.  To this end, \cite{Iosifidis-2014} jointly designed user association and video caching policy by minimizing the user experienced delay while considering different quality requirements for each user. To obtain an optimal trade-off between the content availability and the load balancing, an online algorithm was proposed in \cite{Naveen-2015}. To maximize the system throughput in a coordinated small-cell cellular system, the problem of joint designing of caching, user association, and routing is discussed in \cite{Khreishah-2016}. Considering distinct users have different wireless channels, \cite{Wang-2016} jointly designed the caching and user association policy by minimizing the average delay of small cell UEs in a
heterogeneous network. Jing \textit{et. al.} \cite{jing2019user}, proposed a joint cache delivery and placement strategy by optimizing the user association in different time scales. The line of these works was further
expanded in \cite{Dai-2016} for a cache-enabled Cloud-RAN network
to lessen the backhaul traffic by maximizing a proportional
fairness network utility. However, \cite{Dai-2016} considered a single input single output (SISO) case in which both UEs and BSs
are equipped with a single antenna, and each user is only
connected with one BS.

In practice, the cache placement and the content delivery
(precoding and user association) usually happen in different
timescales. Cache placement usually takes much longer
(e.g., days or hours) than that of content delivery (e.g.,
seconds). Therefore, like \cite{Lau-2013}, we study a mixed-timescale
joint optimization, but in this case for content placement
and content delivery in the cache-enabled cloud radio access
networks to maximize a weighted backhaul-aware network
utility function subject to the peak transmission power and
cache capacity constraints at all BSs. The cache placement
reduces the backhaul consumption and provides more CoMP
opportunities. It is adaptive to the long-term popularity of
data, therefore, the caching strategy should be adaptive to the channel statistics instead of the instantaneous channel
realization in each channel coherent time. In contrast, the
role of the content delivery is to guarantee delivery of a
better average throughput to each user and be adaptive to
instantaneous channel state information.


\subsection{Main Contributions} 
This paper differs from previously studies particularly in its aim to bring a consideration of caching along with user association and resource allocation. We optimize the tradeoff between backhaul reduction and network throughput by maximizing a weighted backhaul-aware network utility function. Furthermore, we consider multiple real-world factors for effective content caching such as popularity distribution, caching placement from the user perspective, and temporal and spatial locality of the content demand, in order to accommodate challenging use cases with strict quality of service requirements. That is why the original problem and, therefore, the three sub-problems in our paper are completely different from the previous works. The main contributions of this paper can be summarized as follows.

\begin{itemize}
  \item For the first time, we define and maximize the network throughput as a function of caching placement strategy, user association policy, precoding vectors, probability that a file is requested by a specific user, and the distance from all connected BSs to this specific user. We consider multiple real-world factors for effective content caching such as popularity distribution, caching placement from the user perspective, and temporal and spatial locality of the content demand, in order to accommodate challenging use cases with strict quality of service requirements.
  \item We introduce the tradeoff between network throughput and backhaul savings by combining user association, MIMO transmit precoding, and cache placement to boost user experience, and identify the interaction between user association, resource allocation and cache placement for a multi-cluster multi-user MEC network.
 
   \item To the best of our knowledge, there is no existing study on jointly optimizing the cache placement and the content delivery. We consider a multi-cluster multi-user MEC network consisting of different users with distinct file preferences, and jointly optimize the mixed-timescale optimization problem of cache placement, user association, and beamforming matrices.
  
  \item We propose an iterative novel algorithm by leveraging the stochastic parallel successive convex approximation (SCA)-based method and the alternating direction method of multipliers (ADMM) to jointly optimize the aforementioned optimization problem. 
  By reducing the complexity, the proposed algorithm is feasible for future wireless big data processing systems.

\end{itemize}

\subsection{Paper Organization and Notations}
The remainder of this paper is organized as follows. In Section II, the system model is described and the main assumptions required for our analysis are introduced. Section III, presents the problem formulation and analysis. Simulation results
are presented in section IV. An overview of the results and concluding remarks are presented in Section V.

\textit{Notation:} Throughout this paper, normal letters are used for scalars. Boldface capital
and lower case letters denote matrices and vectors, respectively. The transposition, the Hermitian
transposition, and the determinant of a complex matrix $\mathbf{A}$ are denoted by $\mathbf{A}^{T}$,
$\mathbf{A}^{H}$ and $\left|\mathbf{A}\right|$, respectively. An $N \times K$ matrix, with ones
on its main diagonal and zeros on its off-diagonal entries, is denoted by $\mathbf{I}_{N \times K}$,
while the identity matrix of size $N$ is simply denoted by $\mathbf{I}_{N}$. An $N \times K$ all-zeros matrix is denoted by $\mathbf{0}_{N \times K}$. The sets of
complex and real numbers are denoted by $\mathbb{C}$ and $\mathbb{R}$, respectively. A circularly
symmetric complex Gaussian random variable (r.v.) is represented by $Z=X+jY \sim \mathcal{CN}(0,\sigma^{2})$,
where $X$ and $Y$ are independent and identically distributed (i.i.d.) normal r.v.'s from
$\mathcal{N}(0,\frac{\sigma^{2}}{2})$. $\mathbb{E}[\cdot]$ represents the expectation operator. The Hadamard product between two matrices $\mathbf{A}$ and $\mathbf{B}$ is symbolized by $\mathbf{A} \odot \mathbf{B}$. 

\section{System Model and Assumptions}
We consider a mobile edge caching network consisting of one central computing unit (cloud), $B$ base stations, and $K$ user equipments  as depicted in Fig. 1. Table \ref{Tb1} summarizes the major notations and symbols used in this paper. The location of the BSs  is modeled by a Poisson Point Process (PPP) with density $\lambda_{B}$ while UEs are distributed around each BS independently and uniformly. We partition the area to $M$ clusters. $|\mathcal{I}_{i}|$ and $|\mathcal{Q}_{i}|$ indicate the number of UEs and BSs in the $i$-th cluster, respectively, where $\mathcal{Q}_{i}\subseteq \{1,2,\ldots, B\}$, $\mathcal{I}_{i}\subseteq \{1,2,\ldots, K\}$, $\mathcal{Q}_{i}\cap \mathcal{Q}_{j} = \varnothing$, $\mathcal{I}_{i}\cap \mathcal{I}_{j} = \varnothing$, $\forall i \neq j, i,j \in \mathcal{M} \triangleq \{1,2,\ldots,M\}$. Let $i_{j}$, $i \in \mathcal{M}$ and $j \in \mathcal{Q}_{i}$, denote the $j$-th BS in the $i$-th cluster and $i_{k}$, $i \in \mathcal{M}$ and $k \in \mathcal{I}_{i}$, indicate the $k$-th user in the $i$-th cluster. BS $i_{j}$ is equipped with $N_{t}^{i_{j}}$ transmit antennas and a cache that stores $s_{i_{j}}$ bits of data while user $i_{k}$ has $N_{r}^{i_{k}}$ receive antennas. The channel (propagation) coefficient between the $i_{j}$ BS and the $i_{k}$ user form channel matrix $\mathbf{G}_{i_{k},i_{j}} = \sqrt{\beta_{i_{k},i_{j}}}\mathbf{H}_{i_{k},i_{j}}\in \mathbb{C}^{N_{r}^{i_{k}} \times N_{t}^{i_{j}}}$ where $\beta_{i_{k},i_{j}}$ is a large-scale fading coefficient that depends upon the shadowing and distance between the corresponding user and BS. The large-scale fading coefficient denoted by $\beta_{i_{k},i_{j}} = \psi_{i_{k},i_{j}}d_{i_{k},i_{j}}^{-\alpha}$, where $d_{i_{k},i_{j}}$ is the distance between the $i_{k}$ user and the $i_{j}$ BS; $\alpha$ is the path-loss exponent; and $\psi_{i_{k},i_{j}}$ is a log-normal random variable, i.e., the quantity $10\log_{10}(\psi_{i_{k},i_{j}})$ is distributed zero-mean Gaussian with a standard deviation of $\sigma_{\text{shadowing}}$. The small-scale fading coefficients, i.e., elements of $\mathbf{H}_{i_{k},i_{j}}$, are modeled as i.i.d. complex Gaussian variables with zero-mean and unit-variance. We further assume a block fading model, where small-scale channels are constant over a few time slots with respect to channel estimation and CSI feedback procedures. Similarly, we assume that large-scale fading coefficients $\beta_{i_{k},i_{j}}$ stay constant during large-scale coherence blocks. The small-scale and large-scale fading coefficients in different coherence blocks are assumed to be independent.
\setlength{\textfloatsep}{10pt}
\begin{figure}
\centering
\includegraphics[scale=0.16]{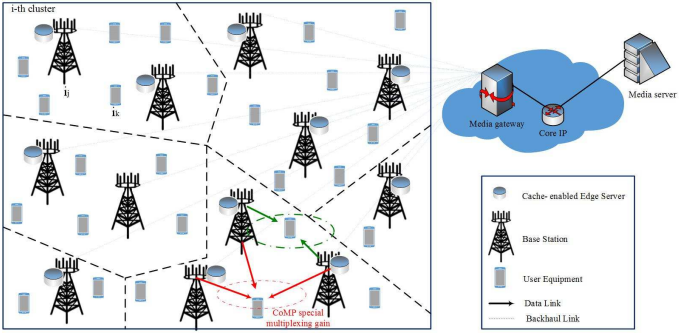}
\caption{Network architecture of mobile edge caching.}
\label{fig: fig1}
\end{figure}

\begin{table}[t]
\captionsetup{font=footnotesize}
  \caption{Nomenclature and Notation Used}\label{Tb1}
  \centering
  \setlength\tabcolsep{0.11cm}
   \scriptsize
   \begin{tabular}{@{} l  l @{}}
    \hline
    Notation &  Description \\ [0.1cm] \hline
    $B$ & Number of BSs  \\ [0.1cm]
    $K$ & Number of UEs \\[0.1cm] 
    $M$ & Number of Clusters \\[0.1cm] 
    $|\mathcal{I}_{i}|$ & Number of UEs in the $i^{\text{th}}$ cluster, \\[0.1cm] 
    $|\mathcal{Q}_{i}|$ & Number of BSs in the $i^{\text{th}}$ cluster, \\[0.1cm] 
    $i_{j}$ & The $j^{\text{th}}$ BS in the $i^{\text{th}}$ cluster \\[0.1cm]
    $i_{k}$ & The $k^{\text{th}}$ UE in the $i^{\text{th}}$ cluster \\[0.1cm]
    $N_{t}^{i_{j}}$ & Number of transmit antennas at the $i_{j}$ BS  \\[0.1cm] 
    $N_{r}^{i_{k}}$ & Number of receive antennas at the $i_{k}$ UE \\[0.1cm] 
    $\mathbf{G}_{i_{k},i_{j}}$ & The channel coefficient between the $i_{j}$ BS and the $i_{k}$ UE\\ [0.1cm]
    $\beta_{i_{k},i_{j}}$ & Large-scale fading coefficients \\ [0.1cm]
    $d_{i_{k},i_{j}}$ & The distance between the $i_{k}$ UE and the $i_{j}$ BS \\[0.1cm]
    $\alpha$ & The path-loss exponent  \\[0.1cm]
    $\psi_{i_{k},i_{j}}$ & A log-normal random variable  \\[0.1cm]
    $\sigma_{\text{shadowing}}$ & The standard deviation of shadowing \\[0.1cm]
    $\mathbf{H}_{i_{k},i_{j}}$ & The small-scale fading coefficients between the $i_{j}$ and the $i_{k}$\\[0.1cm]
    $B_{i_{j}}$ & The capacity of the link connects the $i_{j}$ to the cloud. \\[0.1cm]
    $F$ & Number of data files at the cloud \\[0.1cm]
    $l_{f_{n}}$ & The size of the $f_{n}^{\text{th}}$ file in bits \\[0.1cm]
    $s_{i_{j}}$ & The capacity of the cache at the $i_{j}$ BS \\[0.1cm]
    $q_{i_{k}}$ & Number of requests made by the $i_{k}$ over a given time-interval \\[0.1cm]
    $\pi_{i_{k}}$ & The index of the requested file by the $i_{k}$ user \\[0.1cm]
    $\bar{p}_{i_{k},f_{n}}$ & Probability that the $i_{k}$ UE request file $f_{n}$ \\[0.1cm]
    $p_{i_{j},f_{n}}$ & The popularity distribution of file $n$ observed at the $i_{j}$-th BS \\[0.1cm]
    $\mathbf{D}_{i}$ & The user association matrix in the $i^{\text{th}}$ cluster \\[0.1cm]
    $\mathbf{C}_{i}$ & The cache placement matrix in the $i^{\text{th}}$ cluster \\[0.1cm]
    $\mathbf{C}_{i_{k}}$ &  The caching matrix corresponding to the BSs in $\mathcal{B}_{i_{k}}$\\[0.1cm]
    $\mathcal{B}_{i_{k}}$ & A cooperating set of BSs assigned to the $i_{k}$  \\[0.1cm]
    $\mathcal{K}_{i_{j}}$ & A set of users served by the $i_{j}$  \\[0.1cm]
    $s_{i_{k},i_{j}}$ & The transmit symbol from the $i_{j}$ BS and the $i_{k}$ UE \\[0.1cm]
    $\tilde{p}_{i_{k},i_{j}}$ & The transmit power allocated to the $i_{k}$ UE from the $i_{j}$ BS\\[0.1cm]
    $\mathbf{v}_{i_{k},i_{j}}$ & The unit-norm beamformer from the $i_{j}$ BS to the $i_{k}$ UE \\[0.1cm]
    $P_{i_{j}}^{\max}$ & The transmit power constraint at the $i_{j}$ BS \\[0.1cm]
    \hline
    \end{tabular}
\end{table}
\raggedbottom
Each BS $i_{j}$ is connected to the cloud through a finite-capacity backhaul link $B_{i_{j}}, i \in \mathcal{M}, j \in \mathcal{Q}_{i}$. The cloud has access to the whole data library containing $F$ files, where different contents are independent. Making use of the capacity-limited links restricts the amount of information transfer between the cloud to the BSs. We define $\Pi_{i} = \{\pi_{i_{1}},\ldots,\pi_{i_{|\mathcal{I}_{i}|}}\}$ as the user request profile at the $i$-th cluster, where $\pi_{i_{k}}$ denotes the index of the requested file by the $i_{k}$ user. Users can make random requests from a directory of files $\mathcal{F} = \{f_{1}, f_{2}, \ldots , f_{F}\}$ where each file $f_{n}$ has size $\ell_{f_{n}}$ bits. For the sake of simplicity, we assume that the cache size at any BS is at least large enough to cache any of the files, i.e., $\ell_{f_{n}} \leq s_{i_{j}}$ for all $f_{n} \in \mathcal{F},~ i \in \mathcal{M},~ j \in \mathcal{Q}_{i}$. Moreover, we assume that the $i_{k}$ user makes $q_{i_{k}}$ requests over a given time interval $T$. Therefore, $\mathbf{q}_{i} = [q_{i_{1}}, \dots, q_{i_{|\mathcal{I}_{i}|}}]$ indicates the rates of requests that are made by users in the $i$-th cluster.
We also assume different users in the $i$-th cluster may have different file preferences. Assuming $\bar{p}_{i_{k},f_{n}}$ is the probability that $i_{k}$ user, $i \in \mathcal{M},~k \in \mathcal{I}_{i}$, request file $f_{n} \in \mathcal{F}$, the discrete popularity distribution of files for the users in the $i$-th cluster can be indicated as

\scriptsize
\begin{align}\nonumber
\overline{\mathbf{P}}_{i} =
\begin{bmatrix}
\overline{\mathbf{p}}_{i_{1}} \\
  \vdots  \\
    \overline{\mathbf{p}}_{i_{k}}\\
      \vdots  \\
    \overline{\mathbf{p}}_{i_{|\mathcal{I}_{i}|}}
\end{bmatrix}
=
\begin{bmatrix}
\bar{p}_{i_{1},f_{1}} & \dots  & \bar{p}_{i_{1},f_{F}} \\
\vdots  & \ddots & \vdots \\
\bar{p}_{i_{k},f_{1}}  & \dots  & \bar{p}_{i_{k},f_{F}}\\
\vdots  & \ddots & \vdots \\
\bar{p}_{i_{|\mathcal{I}_{i}|},f_{1}}  & \dots  & \bar{p}_{i_{|\mathcal{I}_{i}|},f_{F}}
\end{bmatrix} \in [0,1]^{|\mathcal{I}_{i}| \times F}
\end{align}
\normalsize
where $\bar{p}_{m,n}$ represents the probability that the $m$-th UE in the $i$-th cluster requests the $n$-th file. Note that the $m$-th row of matrix $\overline{\mathbf{P}}_{i}$ is a stochastic vector that indicates the discrete probability distribution of the $m$-th UE.

Since the file popularity distributions seen at each BS depend on the local file popularities of all connected UEs to the BS \cite{EDebbah-2013}, this matrix will be different from $\overline{\mathbf{P}}_{i}$. The popularity distributions at the BSs in the $i$-th cluster, namely $\mathbf{P}_{i}\in [0,1]^{|\mathcal{Q}_{i}| \times F}$, can be derived as (\ref{Eqn_Pi}) 
\begin{figure*}
\scriptsize
\begin{align}\label{Eqn_Pi}
&\mathbf{P}_{i} =
\begin{bmatrix}
\mathbf{p}_{i_{1}} \\
  \vdots  \\
\mathbf{p}_{i_{|\mathcal{Q}_{i}|}}
\end{bmatrix}
=
\begin{bmatrix}
p_{i_{1},f_{1}} & \dots  & p_{i_{1},f_{F}} \\
p_{i_{2},f_{1}} & \dots  & p_{i_{2},f_{F}} \\
\vdots  & \ddots & \vdots \\
p_{i_{|\mathcal{Q}_{i}|},f_{1}}  & \dots  & p_{i_{|\mathcal{Q}_{i}|},f_{F}}
\end{bmatrix}  =
\begin{bmatrix}
\frac{1}{\mathbf{d}_{i_{1}}\mathbf{q}_{i}^{T}} & 0 & \dots  & 0 \\
0 & \frac{1}{\mathbf{d}_{i_{2}}\mathbf{q}_{i}^{T}} & \dots  & 0 \\
\vdots  & \ddots & \vdots \\
0  & \dots  & 0 & \frac{1}{\mathbf{d}_{i_{|\mathcal{Q}_{i}|}}\mathbf{q}_{i}^{T}}
\end{bmatrix}
\mathbf{D}_{i}
\begin{bmatrix}
q_{i_{1}} & 0 & \dots  & 0 \\
0 & q_{i_{2}} & \dots  & 0 \\
\vdots  & \ddots & \vdots \\
0  & \dots  & 0 & q_{i_{|\mathcal{I}_{i}|}}
\end{bmatrix}
\overline{\mathbf{P}}_{i}
\end{align}
\end{figure*}
where $
p_{i_{j},f_{n}} = \sum_{k =1}^{|\mathcal{I}_{i}|}d_{i_{j},i_{k}}q_{i_{k}}\bar{p}_{i_{k},f_{n}}~/~\sum_{k=1}^{|\mathcal{I}_{i}|}d_{i_{j},i_{k}}q_{i_{k}}
$
denotes the $n$-th file popularity distribution observed at the $j$-th BS in the $i$-th cluster and the denominator is a normalization factor. In practice, by adding up the number of times that the $f_{n}$ file is requested by users, the value of $p_{i_{j},f_{n}}$ can be computed at the cloud. Moreover, since user behavior is correlated with the previously requested data, $p_{i_{j},f_{n}}$ can provide the information regarding the file popularity of the future requests, and thus it helps to efficiently store the files in caches before a request is made. $\mathbf{D}_{i}\in \{0, 1\}^{|\mathcal{Q}_{i}| \times |\mathcal{I}_{i}|}$ denotes the user association matrix in the $i$-th cluster, which depicts the connection between the BSs and UEs in the $i$-th cluster. The user association matrix is structured as

\scriptsize
\begin{align}
\nonumber
\mathbf{D}_{i} =
\begin{bmatrix}
\mathbf{d}_{i_{1}} \\
  \vdots  \\
    \mathbf{d}_{i_{|\mathcal{Q}_{i}|}}
\end{bmatrix}
=
\begin{bmatrix}
d_{i_{1},i_{1}} & \dots  & d_{i_{1},i_{k}} &\dots  & d_{i_{1},i_{|\mathcal{I}_{i}|}} \\
\vdots  & \ddots &\vdots  & \ddots & \vdots \\
d_{i_{|\mathcal{Q}_{i}|},i_{1}}  & \dots  &d_{i_{|\mathcal{Q}_{i}|},i_{k}}  & \dots  & d_{i_{|\mathcal{Q}_{i}|},i_{|\mathcal{I}_{i}|}}
\end{bmatrix}
\end{align}
\normalsize
where $d_{m,n} =  \mathbbm{1}(\tilde{r}_{m,n}\geq r)$. $\tilde{r}_{m,n}$ is the $mn$-th entry of the wireless downlink rate matrix $\tilde{\mathbf{R}}_{i} \in \mathbb{R}^{|\mathcal{Q}_{i}| \times |\mathcal{I}_{i}|}$ in the $i$-th cluster and represents the achievable data rate from BS $m$ to user $n$ while $r$ guarantees a certain quality of service so that the $n$-th user would not connect to the $m$-th BS if the wireless link rate between them was below the threshold $r$.

Since the caching capacity is limited, the aim of designing a cache placement strategy is to store the most popular contents such that the BSs can directly serve the majority of UEs' demands. We define the content placement matrix at the $i$-th cluster as $\mathbf{C}_{i} \in \{0,1\}^{F \times |\mathcal{Q}_{i}|}$, where $\mathbf{C}_{i}(f_{n},i_{j}) = 1$ means the $f_{n}$-th content is stored in the $i_{j}$ BS and $\mathbf{C}_{i}(f_{n},i_{j}) = 0$ represents the opposite. The caching matrix will be different for different user in one cluster. To be specific, the caching matrix corresponding to the BSs that are associated with UE $i_{k}$, $i \in \mathcal{M}$ and $k \in \mathcal{I}_{i}$, can be expressed as
\begin{align}
\mathbf{C}_{i_{k}} = \mathbf{C}_{i}\mathbf{\Lambda}(\mathbf{D}_{i}(:,i_{k}));
\end{align}
where $\mathbf{\Lambda}(\mathbf{x})$ is a diagonal matrix formed from vector $\mathbf{x}$ and $\mathbf{A}(:,n)$ indicates the $n$-th column of the matrix $\mathbf{A}$. Therefore, the sum of all elements in the $n$-th row of $\mathbf{C}_{i_{k}}$ represents the number of BSs who serve the $i_{k}$ user and cache the $f_{n}$-th file. Considering the cache at $i_{j}$ BS can only store $s_{i_{j}}$ bits of data\footnote{We assume that a file is either completely cached or not cached at all in a BS. Of course, partial/coded caching techniques can be envisaged, and these are left for future research.}, the following constraint should be fulfilled at BS $i_{j}$
\begin{align}
\sum_{n = 1}^{F}\mathbf{C}_{i}(f_{n},i_{j})\ell_{f_{n}}\leq s_{i_{j}}, ~~~ \forall i \in \mathcal{M}, ~ j \in \mathcal{Q}_{i}.
\end{align}
Moreover, the $i_{k}$ user is able to download the $f_{n}$-th file from the cache if the following condition is satisfied:
\begin{align}
\sum\limits_{j=1}^{|\mathcal{Q}_{i}|}d_{i_{j},i_{k}}\mathbf{C}_{i}(f_{n},i_{j})> 0;~~ \forall i\in \mathcal{M}, k \in \mathcal{I}_{i},f_{n} \in \mathcal{F},
\end{align}
otherwise, corresponding links may have backhaul cost to transfer the $f_{n}$-th file. As a result, by associating each user with the BSs that cache its requested content, the total backhaul reduction of the aforementioned MEC network can be represented as

\footnotesize
\begin{align}
\nonumber
&\mathbb{E}\{\sum\limits_{i = 1}^{M}\sum\limits_{k =1}^{|\mathcal{I}_{i}|}\sum\limits_{j=1}^{|\mathcal{Q}_{i}|}d_{i_{j},i_{k}}\mathbf{C}_{i}(f_{n},i_{j})\} = \\ \nonumber &
\sum\limits_{i = 1}^{M}\sum\limits_{k =1}^{|\mathcal{I}_{i}|}\sum\limits_{n = 1}^{F}\sum\limits_{j=1}^{|\mathcal{Q}_{i}|}d_{i_{j},i_{k}}\bar{p}_{i_{k},f_{n}}\mathbf{C}_{i}(f_{n},i_{j}) = \sum\limits_{i = 1}^{M}\text{Tr}(\overline{\mathbf{P}}_{i}\mathbf{C}_{i}\mathbf{D}_{i}),
\end{align}\normalsize
where expectation is with respect to the random user requests.

Furthermore, we assume that in the underlying MEC network, a cluster of cooperative BSs serves each UE. To be specific, a cooperating set $\mathcal{B}_{i_{k}}$ is assigned to $i_{k}$ user where $\mathcal{B}_{i_{k}} = \{i_{j}|d_{i_{j},i_{k}} = 1\}\subseteq \mathcal{Q}_{i}$ is formed by aggregating all the BSs that have knowledge of channels $\mathbf{H}_{i_{k},i_{j}},~ i_{j} \in \mathcal{B}_{i_{k}}$, and have access to $i_{k}$ user's message. They may also jointly encode the message intended for this user in their transmission \cite{Ng-2010}. Note that, due to the shadowing effect, the $|\mathcal{B}_{i_{k}}|$ strongest BSs under coordination are not necessarily the $|\mathcal{B}_{i_{k}}|$ nearest BSs, where $|\mathcal{B}_{i_{k}}|=\|\mathbf{D}_{i}(:,i_{k})\|_{0}$ denote the set's cardinality. Since each BS may get involved in transmission to more than one UE, the cooperating set of different users may overlap. We designate the set of users served by the $i_{j}$ BS by $\mathcal{K}_{i_{j}}\triangleq \{i_{k}| i_{j} \in \mathcal{B}_{i_{k}}\}\subseteq  \mathcal{I}_{i}$. It is worth nothing that since the number of users that a BS can support in a specific frequency and time resource block is limited by the number of BS antennas, it is assumed that $\mathcal{K}_{i_{j}} \leq N_{t}^{i_{j}}$. Moreover, similar to \cite{MoslehIA-2016,Mosleh-2015}, we assume each co-scheduled user operating in the MU-MIMO mode only receives one spatial stream (rank 1 transmission) as specified by the Rel-10 LTE-Advanced standard \cite[Chapter 11]{Sesia-2011}. BS $i_{j}$ plans to communicate a symbol vector $\mathbf{s}_{i_{j}} = [s_{i_{1},i_{j}}, \ldots, s_{i_{|\mathcal{K}_{i_{j}}|},i_{j}}]^{T} \in \mathbb{C}^{|\mathcal{K}_{i_{j}}|}$ to its associated receivers, where $s_{i_{k},i_{j}}$ is the transmit symbol from the $i_{j}$ BS to the $i_{k} \in \mathcal{K}_{i_{j}}$ receiver with unit power of $\mathbb{E}\{|s_{i_{k},i_{j}} |^{2}\} = 1$ and $|\mathcal{K}_{i_{j}}|$ denote the set's cardinality.
\begin{remark}
The main idea behind defining $M$ and $\mathcal{B}_{i_{k}}$ is that the demand for content shows variations both across time and space and, in this paper, we considered both temporal and spatial locality of the content demand. To be specific, it is crucial to consider the user location since the demand may differentiate from one geographical area to another. This space variation of the content request, referred to as spatial locality, is why we define $M$ clusters, where $M$ can be any arbitrary number. On the other hand, the demand may vary from time to time. This temporal variation of the content request, referred to as time locality, is the reason that we defined the $\mathcal{B}_{i_{k}}$ cluster associated with user $i_{k}$.
\end{remark}

Prior to transmitting, the $i_{j}$ BS linearly precodes its symbol vector $\mathbf{x}_{i_{j}} = \sum_{i_{k} \in \mathcal{K}_{i_{j}}}\sqrt{\tilde{p}_{i_{k},i_{j}}}\mathbf{v}_{i_{k},i_{j}}s_{i_{k},i_{j}}$ where $\tilde{p}_{i_{k},i_{j}}$ stands for the transmit power allocated to user $i_{k}$ from the $i_{j}$ BS and $\mathbf{v}_{i_{k},i_{j}}$ denotes the unit-norm beamformer that BS $i_{j}$ uses to transmit the signal $s_{i_{k},i_{j}}$ to receiver $i_{k}$. Each BS $i_{j}$ is under a transmit power constraint of $P_{i_{j}}^{\max}$ and so, the transmit power at the $i_{j}$ BS is computed as $\tilde{P}_{i_{j}} = \mathbb{E}\{\|\mathbf{x}_{i_{j}}\|^2\}= \sum_{i_{k} \in \mathcal{K}_{i_{j}}}\tilde{p}_{i_{k},i_{j}}\mathbf{v}_{i_{k},i_{j}}^{H}\mathbf{v}_{i_{k},i_{j}} = \sum_{i_{k} \in \mathcal{K}_{i_{j}}}\tilde{p}_{i_{k},i_{j}}\leq P_{i_{j}}^{\max}$ \cite{MoslehIA-2016, Mosleh-2015}. Under our assumptions, when user $i_{k}$ requests file $f_{n}$ which is available at the cache of the BSs in $\mathcal{B}_{i_{k}}$, the received signals from these BSs are combined coherently using coordinated joint transmission. Since there are $|\mathcal{B}_{i_{k}}|$ BSs participating in the cooperating data transmission to user $i_{k}$, at the same frequency and time, we denote $s_{i_{k}}\in \mathbb{C}$ as the complex data symbol for the $i_{k}$ UE, where $i \in \mathcal{M}$ and $k \in \mathcal{I}_{i}$. Consequently, throughout each symbol duration time, the cooperating BSs transmit the same symbol $s_{i_{k}}$ and the expected received signal vector $\mathbf{y}_{i_{k},f_{n}}\in \mathbb{C}^{N_{r}^{i_{k}} \times 1}$ at the $k$-th user in the $i$-th cluster, when user $i_{k}$ requested file $f_{n}$, can be written as 

\footnotesize
\begin{align}\nonumber
\mathbf{y}_{i_{k},f_{n}} &=  \sum\limits_{j = 1}^{|\mathcal{Q}_{i}|}\mathbf{C}_{i_{k}}(f_{n},i_{j})\sqrt{\tilde{p}_{i_{k},i_{j}}}\sqrt{\beta_{i_{k},i_{j}}}\mathbf{H}_{i_{k},i_{j}}\mathbf{v}_{i_{k},i_{j}}s_{i_{k}}+ \mathbf{n}_{i_{k}} 
\end{align}
\begin{align}\nonumber
&+
\sum\limits_{n=1}^{F}\sum\limits_{\substack{m=1 \\ m\neq k}}^{|\mathcal{I}_{i}|}\sum\limits_{j = 1}^{|\mathcal{Q}_{i}|}\mathbf{C}_{i_{m}}(f_{n},i_{j})\sqrt{\tilde{p}_{i_{m},i_{j}}}\sqrt{\beta_{i_{k},i_{j}}} \times \mathbf{H}_{i_{k},i_{j}}\mathbf{v}_{i_{m},i_{j}}s_{i_{m}} 
\\ \nonumber &+
\sum\limits_{n=1}^{F}\sum\limits_{\substack{q=1 \\ q\neq i}}^{M}\sum\limits_{\ell = 1}^{|\mathcal{I}_{q}|}\sum\limits_{j =1}^{|\mathcal{Q}_{q}|}
\mathbf{C}_{q_{\ell}}(f_{n},q_{j})\sqrt{\tilde{p}_{q_{\ell},q_{j}}}\sqrt{\beta_{i_{k},q_{j}}}\mathbf{H}_{i_{k},q_{j}}\mathbf{v}_{q_{\ell},q_{j}}s_{q_{\ell}},
\end{align}
\normalsize
where the first term on the right-hand side represents the received useful signal, the third and fourth terms represent the intra- and inter-cluster interference respectively, and $\mathbf{n}_{i_{k}}\sim \mathcal{CN}(0, \sigma_{i_{k}}^{2}\mathbf{I})$ is the additive white Gaussian noise (AWGN) at the $i_{k}$ UE. We assumed that the signals for different users are independent from each other. In  this paper, we  treat interference as  noise and consider a linear receive beamforming strategy so that the estimated signal is given by $\hat{s}_{i_{k}} = \mathbf{u}_{i_{k}}^{H}\mathbf{y}_{i_{k},f_{n}}$. Indeed, each receiver $i_{k} \in \mathcal{K}_{i_{j}}$, linearly processes the received signal to obtain $\mathbf{u}_{i_{k}}^{H}\mathbf{y}_{i_{k},f_{n}}$ where $\mathbf{u}_{i_{k}} \in \mathbb{C}^{N_{r}^{i_{k}}}$ denotes the unit-norm post processing filter at receiver $i_{k}$, i.e., $\|\mathbf{u}_{i_{k}}\|^{2} = 1$. 

As mentioned above, the received interference at the $i_{k}$-th UE is the summation of the intra-cluster and inter-cluster interferences. While the former is the interference experienced by the $i_{k}$ UE from all BSs inside the $i$-th cluster, i.e.,

\small
\begin{align}
\nonumber
I_{i_{k},f_{n}}^{\text{intra}} = \sum\limits_{n=1}^{F}\sum\limits_{\substack{m=1 \\ m\neq k}}^{|\mathcal{I}_{i}|}\sum\limits_{j=1}^{|\mathcal{Q}_{i}|}&\mathbf{C}_{i_{m}}(f_{n},i_{j})\tilde{p}_{i_{m},i_{j}}\beta_{i_{k},i_{j}}|\mathbf{v}_{i_{m},i_{j}}^{H}\mathbf{H}_{i_{k},i_{j}}^{H}\mathbf{u}_{i_{k}}|^2, 
\end{align}
\normalsize
the latter is the received interference from all the BSs outside the $i$-th cluster and can be represented as

\small
\begin{align}
\nonumber
I_{i_{k},f_{n}}^{\text{inter}} = \sum\limits_{n=1}^{F}\sum\limits_{\substack{q=1 \\ q\neq i}}^{M}\sum\limits_{\ell=1}^{|\mathcal{I}_{q}|}\sum\limits_{j =1}^{|\mathcal{Q}_{q}|}&\mathbf{C}_{q_{\ell}}(f_{n},q_{j})\tilde{p}_{q_{\ell},q_{j}}\beta_{i_{k},q_{j}}|\mathbf{v}_{q_{\ell},q_{j}}^{H}\mathbf{H}_{i_{k},q_{j}}^{H}\mathbf{u}_{i_{k}}|^2.
\end{align}
\normalsize

Therefore, the expected SINR at the $k$-th user in the $i$-th cluster, when user $i_{k}$ requests file $f_{n}$, is a function of transmit power, cache placement matrix, user association matrix, probability of file being requested, and distance from all connected BSs to $i_{k}$ user and can be written as Eqn. (\ref{equation11}), where $\mathbf{C}_{q_{\ell}}(f_{n},q_{j}) = \mathbf{C}_{q}(f_{n},q_{j})\mathbf{D}_{q}(q_{j},q_{\ell})$. Therefore the average transmission rate over channel realizations for user $i_{k}$ can be written as $R_{i_{k},f_{n}} = \mathbb{E}[\log(1+ \text{SINR}_{i_{k},f_{n}})]$ and the total average throughput of this user can be formulated as
\begin{figure*}[b]
\small
\begin{align}\label{equation11}
\text{SINR}_{i_{k},f_{n}}=\frac{\sum\limits_{j = 1}^{|\mathcal{Q}_{i}|}\mathbf{C}_{i_{k}}(f_{n},i_{j})\tilde{p}_{i_{k},i_{j}}\beta_{i_{k},i_{j}} |\mathbf{u}_{i_{k}}^{H}
\mathbf{H}_{i_{k},i_{j}}\mathbf{v}_{i_{k},i_{j}}|^{2} }{\mathbf{u}_{i_{k}}^{H}\big(  \sum\limits_{n=1}^{F}\sum\limits_{(\ell, q)\neq (k,i)}\sum\limits_{j =1}^{|\mathcal{Q}_{q}|}\mathbf{C}_{q_{\ell}}(f_{n},q_{j})\tilde{p}_{q_{\ell},q_{j}}\beta_{i_{k},q_{j}}\mathbf{H}_{i_{k},q_{j}}\mathbf{v}_{q_{\ell},q_{j}}\mathbf{v}_{q_{\ell},q_{j}}^{H}\mathbf{H}_{i_{k},q_{j}}^{H} + \sigma_{i_{k}}\mathbf{I}\big)\mathbf{u}_{i_{k}}}~,
\end{align}
\end{figure*}
\begin{align}\nonumber
\small
R_{i_{k}} = \mathbb{E}\{R_{i_{k},f_{n}}\} = \sum\limits_{n=1}^{F}\bar{p}_{i_{k},f_{n}}R_{i_{k},f_{n}}, ~~~ \forall i\in \mathcal{M},~k \in \mathcal{I}_{i}
\end{align}
\normalsize
where expectation is with respect to the random user requests. Note that, the network throughput does not depend on the cache placement matrix $(\mathbf{C}_{i})$ but depends on the caching matrix corresponding to the BSs that are associated with a specific user $(\mathbf{C}_{i_{k} })$. To be specific, no matter the content is cached or not cached in the BS, the user needs to communicate with its associated BSs. The distribution of the UEs' content demands follows a Zipf-like distribution $P_{\mathcal{F}}(i)$ given as $P_{\mathcal{F}}(i) = 1/i^{\gamma}\sum_{i=1}^{F}i^{-\gamma}, i \in \mathcal{F}$, where $\gamma$ models the skewness of the popularity profile \cite{Breslau-1999,Hefeeda-2008}. Depending on both the BSs deployment strategies and the UEs' behavior, $\gamma$ takes different values. The popularity is uniformly distributed over content files for lower values of $\gamma$, meaning UEs have more distinct interests. As $\gamma$ grows, the popularity becomes more skewed towards the most popular files, which means UEs have very similar file interests and a small subset of files are more desired than the rest.

\section{Problem Formulation and Analysis}
In this section, the problem of interest is a joint optimization of the content placement and content delivery (precoding and user association). We formulate a mixed-timescale optimization problem which maximizes the trade-off between the backhaul savings and the network throughput. The motivation behind maximizing the trade-off between the backhaul savings and the network throughput is providing a new technique in MEC that can support both low latency (in a low-cost manner) and high throughput requirements for the upcoming 5G wireless network. For maximizing the backhaul reduction, each UE should be associated with a BS (cluster of BSs) that caches the largest amount of its desired contents. However, such a BS might be a long way away. On the other hand, in order to maximize the network throughput, each UE should take the BSs' load into account and accordingly associate with a BS that provides it a reasonably high SINR. Nonetheless, such a BS may not store the user's desired contents. Considering such a tradeoff into account, in this paper the problem of maximizing the user throughput and backhaul savings with respect to the precoding matrix, cache placement matrix, and user association matrix, is formulated subject to peak transmission power, cache capacity, and backhaul capacity constraints.

Let $\mathbf{\mathcal{V}} = \big\{\mathbf{v}_{i_{k},i_{j}}(\Pi_{i},\{\mathbf{H}_{i_{k},i_{j}}\}_{j=1}^{|\mathcal{Q}_{i}|}): \forall i,k,j\big\}$ and $\mathbf{\mathcal{D}} = \big\{\mathbf{D}_{i}(\Pi_{i},\{\mathbf{H}_{i_{k},i_{j}}\}_{j=1}^{|\mathcal{Q}_{i}|}):\forall i,k,j\big\}$ denote all the beamforming vectors and user-association matrices for all user request profiles $\{\Pi_{i}\}_{i=1}^{M}$ and instantaneous CSI matrices $\{\mathbf{H}_{i_{k},i_{j}}\}_{j=1}^{|\mathcal{Q}_{i}|}$, respectively. Then for a given set of optimization variables $(\mathbf{\mathcal{V}},\mathbf{\mathcal{D}},\mathbf{C}_{i})$ and user request profiles $\{\Pi_{i}\}_{i=1}^{M}$, the average utility function of our optimization problem can be expressed as $\sum_{i = 1}^{M}\sum_{k=1}^{|\mathcal{I}_{i}|} \mathbb{E}\big[\lambda R_{i_{k},f_{n}} + (1-\lambda)\sum_{j=1}^{|\mathcal{Q}_{i}|}d_{i_{j},i_{k}}\mathbf{C}_{i}(f_{n},i_{j})\big|\Pi_{i}\big]$ where the expectation is taken with respect to both the CSI and the distribution of user requests, and the parameter controlling the tradeoff between network throughput and backhaul savings is denoted by $\lambda\in (0,1)$ so that by adjusting $\lambda$ we can emphasize one term over the other. The optimization variables are partitioned into short-term (user association and beamforming) and long-term (content placement) variables. While the latter is adaptive to the popularity of data and the channel statistics, the former is adaptive to the instantaneous CSI. The following feasible sets are defined for the cache placement $\mathbf{C}_{i}$, the user association $\mathbf{D}_{i}$, and the beamforming vectors $\mathbf{v}_{i_{k},i_{j}}$ for all $i \in \mathcal{M}, j \in \mathcal{Q}_{i}, k \in \mathcal{I}_{i}$ as follows\footnote{It is worth mentioning that although the capacity capability of cloud is much higher than that of caching, but it is by no means that the cloud capability is unlimited in the cloud. Investigating the effect of this constraint in a multi-cluster multi-user cache-enabled CRAN would be an important topic for future work.} 

\footnotesize
\begin{align}\nonumber
\mathcal{S}_{\mathbf{C}} &= \{\mathbf{C}_{i}:~\mathbf{C}_{i}(f_{n},i_{j})\in\{0,1\}~,~\sum\limits_{n = 1}^{F}\mathbf{C}_{i}(f_{n},i_{j})\ell_{f_{n}}\leq s_{i_{j}}\},\\ \nonumber 
\mathcal{S}_{\mathbf{D}} &= \{\mathbf{D}_{i}:~\mathbf{D}_{i}(i_{j},i_{k})\in\{0,1\}\},\\ \nonumber 
\mathcal{S}_{\mathbf{v}} &= \{\mathbf{v}_{i_{k},i_{j}}:~\sum\limits_{k=1}^{|\mathcal{I}_{i}|} \text{Tr}(\mathbf{v}_{i_{k},i_{j}}^{H}\tilde{p}_{i_{k},i_{j}}\mathbf{v}_{i_{k},i_{j}})\leq P_{i_{j}}^{\max}\}.
\end{align}
\normalsize




Then the joint content placement and content delivery problem $\mathcal{P}$ is formulated as follows\footnote{Note that we are considering a particular realization of the PPP, i.e., $B$ is a sample of a Poisson random variable.}:

\small
\begin{equation}
\begin{aligned}\label{equation20}
& \underset{\mathbf{C},\mathbf{D},\mathbf{v}}{\text{maximize}}
& & \sum\limits_{i = 1}^{M}\sum\limits_{k=1}^{|\mathcal{I}_{i}|} \mathbb{E}\big[\lambda R_{i_{k},f_{n}} + (1-\lambda)\sum\limits_{j=1}^{|\mathcal{Q}_{i}|}d_{i_{j},i_{k}}\mathbf{C}_{i}(f_{n},i_{j})\big|\Pi_{i}\big] \\
& \text{subject to}
& &  \mathbf{C}_{i}\in\mathcal{S}_{\mathbf{C}}~,~\mathbf{v}_{i_{k}, i_{j}}\in \mathcal{S}_{\mathbf{v}}~,~ \mathbf{D}_{i} \in \mathcal{S}_{\mathbf{D}}\\
& & &\sum\limits_{i_{k} \in \mathcal{K}_{i_{j}}}d_{i_{j},i_{k}}R_{i_{k}}\leq B_{i_{j}}, ~ \forall i \in \mathcal{M}, ~ j \in \mathcal{Q}_{i}
\end{aligned}
\end{equation}
\normalsize
where $\sum_{i_{k} \in \mathcal{K}_{i_{j}}}d_{i_{j},i_{k}}R_{i_{k}}$ denotes the $i_{j}$ BS's backhaul consumption. The objective function can be expressed in a more compact form as $\sum_{i = 1}^{M}\mathbb{E}[\text{Tr}\big(\lambda\mathbf{E}_{i}(\overline{\mathbf{P}}_{i}\odot \mathbf{R}_{i})+ (1-\lambda)\overline{\mathbf{P}}_{i}\mathbf{C}_{i}\mathbf{D}_{i}\big)]$ in which $\mathbf{E}_{i}$ is a $F \times |\mathcal{I}_{i}|$ matrix full of $1$'s, $\mathbf{R}_{i}$ is a $|\mathcal{I}_{i}| \times F$ rate matrix so that its $(i_{k}-f_{n})$-th element is equal to $R_{i_{k},f_{n}}$, and the expectation is taken with respect to the distribution of user requests.

Due to the coupled interference among users and the limited-backhaul capacity constraint, the optimization problem (\ref{equation20}) is non-convex. Moreover, the entries of user association and cache placement matrices take binary values $0$ and $1$, thus the optimization problem falls into a mixed integer nonlinear programming (MINLP) which is NP-hard in general \cite{Kleinberg-2005} and non-tractable in practice. Since it is highly unlikely to compute a globally optimal solution in polynomial time, our goal is to obtain a tractable near-optimal solution by developing effective suboptimal algorithms. By utilizing the timescale separations of the optimization variables and making use of the fact that all constraints are separable, we divide the original optimization problem into three subproblems and propose an iterative algorithm that at each time maximizes the objective function with respect to one variable while assuming the other two variables are given. Therefore, each of these subproblems can be relaxed to a convex problem so that it can be solved efficiently. The mixed-timescale joint optimization of content delivery and content placement can be decomposed to the following sub-problems:

\subsection{\textbf{Short-term Content Delivery}}
As mentioned before, the content placement usually takes much longer than the content delivery. Therefore, in this subsection, it is assumed that the user request and the content placement matrix $\mathbf{C}_{i}$ is fixed and given. Therefore, we pay attention to the joint optimization of beamforming design and user-association (with respect to the user request). We further decouple the joint optimization problem in two stages. At the first stage, we associated each user with a cluster of BSs and at the second stage, assuming the user-association is fixed and given, we design the beamformers. The proposed algorithm is described as follows:

\textbf{$\diamond$ ~User-association Stage:}
~Assuming caching policy and beamforming vectors are given, the user association problem can be simplified as

\small
  \begin{equation}
\begin{aligned}\label{equation36}
\mathcal{P}_{1}:~& \underset{\mathbf{D}_{i}}{\text{maximize}}
& & \sum\limits_{i = 1}^{M}\sum\limits_{k =1}^{|\mathcal{I}_{i}|}\sum\limits_{j=1}^{|\mathcal{Q}_{i}|}d_{i_{j},i_{k}}\mu_{i_{j},i_{k}} \\\nonumber
& \text{subject to}
&& d_{i_{j},i_{k}}(1 - d_{i_{j},i_{k}}) = 0, \forall i \in \mathcal{M}, j \in \mathcal{Q}_{i}, k \in \mathcal{I}_{i}\\ \nonumber
& & &\sum\limits_{i_{k} \in \mathcal{K}_{i_{j}}}d_{i_{j},i_{k}}R_{i_{k}}\leq B_{i_{j}}, ~ \forall i \in \mathcal{M}, ~ j \in \mathcal{Q}_{i}
\end{aligned}
\end{equation}
\normalsize
where $\mu_{i_{j},i_{k}} = \sum_{n = 1}^{F}\bar{p}_{i_{k},f_{n}}\mathbf{C}_{i}(f_{n},i_{j})$ represents the amount of backhaul savings by associating the $i_{k}$-th UE with the $i_{j}$-th BS. Due to the fact that the entries of $\mathbf{D}_{i}$ take binary values $0$ and $1$, the optimization problem is a mixed integer optimization over the user association. In order to solve this optimization problem and inspired by the idea used in \cite{Shen-2014}, this paper adopts a new approach called dual analysis. The main idea behind this method is to answer how the optimal value can be deduced from the constraints. This method is used in \cite{Shen-2014} and \cite{Dai-2016} to find a solution to the user association problem in heterogeneous cellular networks under the proportional fairness criterion. Using this method, the optimization problem can be easily decoupled among the clusters and the solution can be expressed as:
\begin{align}\label{equation37}
\small
d^{\ast}_{i_{j},i_{k}} =
\begin{cases}
    1   ,    & \quad \text{if } \mu_{i_{j},i_{k}}> 0\\
    0   ,    & \quad \text{Otherwise} \\
\end{cases}
\end{align}
\normalsize
which shows that taking caching placement and thus backhaul reduction into account can be viewed as an additional incentive for UEs to associate with a BS.

  \textbf{$\diamond$~Beamforming Stage}: With fixed user association and request, the problem of designing beamforming vectors $\mathbf{v}_{i_{k}}$ and $\mathbf{u}_{i_{k}}$, $i \in \mathcal{M}, ~ j \in \mathcal{Q}_{i}$, can be written as  
  \begin{equation}
  \small
\begin{aligned}\label{equation23}
\mathcal{P}_{2}:~& \underset{\substack{\mathbf{v}_{i_{k},i_{j}} \\ \mathbf{u}_{i_{k},i_{j}}}}{\text{maximize}}
& & \sum_{i = 1}^{M}\sum_{k=1}^{|\mathcal{I}_{i}|} \log(1 + \text{SINR}_{i_{k},f_{n}})\\
& \text{subject to}
& &  \sum_{k=1}^{|\mathcal{I}_{i}|}\tilde{p}_{i_{k},i_{j}} \parallel\mathbf{v}_{i_{k},i_{j}}\parallel^{2}\leq P_{i_{j}},\\
& & &\sum\limits_{i_{k} \in \mathcal{K}_{i_{j}}}d_{i_{j},i_{k}}\log(1+ \text{SINR}_{i_{k},f_{n}})\leq B_{i_{j}},
\end{aligned}
\end{equation}
\normalsize
where the optimal short-term bemformers will be calculated with respect to the given user request and thus they do not depend on the file request probability. In order to suit our system model we applied the weighted minimum mean-square error (WMMSE) framework \cite{Shi-2011} into the above optimization problem and modified Proposition 3.2. in \cite{Ma-2016}. In this way, each user can connect to a cluster of BSs instead of being served by only one BS. Thereby, (\ref{equation23}) can be effectively rewritten as

\small
\begin{equation}
\begin{aligned}\label{equation24}
& \underset{\substack{\mathbf{v}_{i_{k},i_{j}} \\ \mathbf{u}_{i_{k},i_{j}}\\ w_{i_{k}}}}{\text{maximize}}
& & \sum\limits_{i = 1}^{M}\sum\limits_{k=1}^{|\mathcal{I}_{i}|} \big(\log(w_{i_{k}})- w_{i_{k}}\varepsilon_{i_{k}}+1\big)\\
& \text{subject to}
& &  \sum\limits_{k=1}^{|\mathcal{I}_{i}|}\tilde{p}_{i_{k},i_{j}} \parallel\mathbf{v}_{i_{k},i_{j}}\parallel^{2}\leq P_{i_{j}},\\
& & &\sum\limits_{i_{k} \in \mathcal{K}_{i_{j}}}\varrho_{i_{k},i_{j}}d_{i_{j},i_{k}}\hat{R}_{i_{k}}||\mathbf{v}_{i_{k},i_{j}}||^{2}\leq B_{i_{j}}, 
\end{aligned}
\end{equation}
\normalsize
where $\{w_{i_{k}}\}$ are the weights variable introduced by WMMSE framework and $\{\varepsilon_{i_{k}}\}$ are the mean square estimation errors which are defined by

\footnotesize
\begin{align}\nonumber
\varepsilon_{i_{k}}&\triangleq |1 - \sum\limits_{j = 1}^{|\mathcal{Q}_{i}|}\mathbf{C}_{i_{k}}(f_{n},i_{j})\tilde{p}_{i_{k},i_{j}}\beta_{i_{k},i_{j}} \mathbf{u}_{i_{k}}^{H}
\mathbf{H}_{i_{k},i_{j}}\mathbf{v}_{i_{k},i_{j}}|^{2} +\sigma_{i_{k}}\|\mathbf{u}_{i_{k}}\|^{2}\\ \nonumber &
+ \sum\limits_{(\ell, q)\neq (k,i)}\sum\limits_{j =1}^{|\mathcal{Q}_{q}|}\mathbf{C}_{q_{\ell}}(f_{n},q_{j})\tilde{p}_{q_{\ell},q_{j}}\beta_{i_{k},q_{j}}|\mathbf{u}_{i_{k}}^{H}\mathbf{H}_{i_{k},q_{j}}\mathbf{v}_{q_{\ell},q_{j}}|^{2},
\end{align}
\normalsize
and $\varrho_{i_{k},i_{j}}$ and $\hat{R}_{i_{k}}$ are the iteratively updated weights with a small positive regularization factor and the achievable data rate calculated from the previous iteration as in \cite{Binbin-2015}, respectively.

The objective function of (\ref{equation24}) is convex with respect to each of the optimization variables $\mathbf{v}_{i_{k},i_{j}}$, $\mathbf{u}_{i_{k},i_{j}}$, and $w_{i_{k}}$, which enables us to employ the block coordinate descent method to solve it \cite{Shi-2011}. To be specific, we maximize the cost function of (\ref{equation24}) by updating one of three variables $\mathbf{v}_{i_{k},i_{j}}$, $\mathbf{u}_{i_{k},i_{j}}$, and $w_{i_{k}}$, while assuming the rest are given. In particular, we iteratively run the following steps.

\noindent$\bullet$~Initializing all the transmit beamformers $\mathbf{v}_{i_{k},i_{j}}$'s, $\forall i, k, j$, and minimizing the weighted sum-MSE leads us to the MMSE receiver $\mathbf{u}_{i_{k}}$ as follows

 \small
  \begin{align}\label{equation26}
  \mathbf{u}_{i_{k}}^{\text{mmse}} = \mathbf{J}_{i_{k}}^{-1} \mathbf{H}_{i_{k},i_{j}}\mathbf{v}_{i_{k},i_{j}}
  \end{align}
  \normalsize
  where
  \footnotesize
  \begin{align}\nonumber
  \mathbf{J}_{i_{k}} = \sum\limits_{q=1}^{M}\sum\limits_{\ell = 1}^{|\mathcal{I}_{q}|}\sum\limits_{j =1}^{|\mathcal{Q}_{q}|}
\mathbf{C}_{q_{\ell}}(f_{n},q_{j})\tilde{p}_{q_{\ell},q_{j}}\beta_{i_{k},q_{j}}
||\mathbf{H}_{i_{k},q_{j}}\mathbf{v}_{q_{\ell},q_{j}}||^{2}+ \sigma_{i_{k}}^{2}\mathbf{I}
\end{align}
\normalsize
 is the covariance matrix of the total received signal at UE $i_{k}$.
\raggedbottom

\noindent$\bullet$~ By fixing all $\mathbf{u}_{i_{k}}$'s and $\mathbf{v}_{i_{k},i_{j}}$'s, $\forall i, j, k$, the weights, for all $i$ and $k$, can be updated as follows

      \small
  \begin{align}\label{equation27}
  w_{i_{k}} = (1-\sum\limits_{j = 1}^{|\mathcal{Q}_{i}|}\mathbf{C}_{i_{k}}(f_{n},i_{j})\tilde{p}_{i_{k},i_{j}}\beta_{i_{k},i_{j}} \mathbf{u}_{i_{k}}^{H}
\mathbf{H}_{i_{k},i_{j}}\mathbf{v}_{i_{k},i_{j}})^{-1}
  \end{align}\normalsize
\noindent$\bullet$~ By fixing all $w_{i_{k}}$'s and $\mathbf{u}_{i_{k}}$'s, the transmit beamformers can be calculated using the following optimization problem

  \small
  \begin{equation}
\begin{aligned}\label{equation28}
& \underset{\mathbf{v}_{i_{k},i_{j}}}{\text{minimize}}
& & \sum\limits_{i = 1}^{M}\sum\limits_{k=1}^{|\mathcal{I}_{i}|}  w_{i_{k}}\varepsilon_{i_{k}}\\
& \text{subject to}
& &  \sum\limits_{k=1}^{|\mathcal{I}_{i}|}\tilde{p}_{i_{k},i_{j}} \parallel\mathbf{v}_{i_{k},i_{j}}\parallel^{2}\leq P_{i_{j}}~,\\ &&&\sum\limits_{i_{k} \in \mathcal{K}_{i_{j}}}d_{i_{j},i_{k}}\hat{R}_{i_{k}}||\mathbf{v}_{i_{k},i_{j}}||^{2}\leq B_{i_{j}}
\end{aligned}
\end{equation}
\normalsize
Problem (\ref{equation28}) is convex and (\ref{equation26}) and (\ref{equation27}) can be locally implemented at the users. Therefore, we solve the problem (\ref{equation28}) in a distributed manner based on the alternating direction method of multipliers (ADMM) \cite{Parikh-2011}. 
In what follows, it is shown that by exchanging a fair amount of information between UEs and BSs the ADMM can be applied in a distributed fashion to solve optimization problem (\ref{equation28}).
In order to achieve a distributed implementation of the ADMM in the aforementioned MEC network, the following assumptions are made (similar to \cite{Shi-2011} and \cite{Ma-2016}). We assume that each BS $j \in Q_{i}$ knows $\mathbf{H}_{i_{k},i_{j}}$ for all $i_{k}$ user in cluster $\mathcal{B}_{i_{k}}$ and each user $i_{k}$ can estimate the interference plus noise covariance matrix. Under these assumptions, the ADMM can be applied distributively. Note that, in order to identify the beamforming vectors in a distributed fashion, our aim here is to have a same form as the one in \cite{Parikh-2011}. To do so, we introduce auxiliary variables $\{\mathbf{x}_{i_{k},i_{j}}\}$ and $\{X_{i_{k}}^{i_{k},i_{j}}\}$ and rewrite problem (\ref{equation28}) as (13).

\begin{figure*}[b]
\footnotesize
    \begin{mini}
      {\substack{\mathbf{v}_{i_{k},i_{j}},\mathbf{x}_{i_{k},i_{j}}, X_{i_{k}}^{i_{k},i_{j}}}}{\sum\limits_{i = 1}^{M} \sum\limits_{k =1}^{|\mathcal{I}_{i}|}\big(|\sqrt{ w_{i_{k}}} - \sum\limits_{j = 1}^{|\mathcal{Q}_{i}|}\bar{p}_{i_{k},f_{n}}^{2}\mathbf{C}_{i_{k}}(f_{n},i_{j})\tilde{p}_{i_{k},i_{j}}\beta_{i_{k},i_{j}} X_{i_{k}}^{i_{k},i_{j}}|^{2}
\sum\limits_{(\ell, q)\neq (k,i)}\sum\limits_{j =1}^{|\mathcal{Q}_{q}|}\mathbf{C}_{q_{\ell}}(f_{n},q_{j})\tilde{p}_{q_{\ell},q_{j}}\beta_{i_{k},q_{j}}|X_{i_{k}}^{q_{\ell},q_{j}}|^{2}\big)}{}{}
      \addConstraint{\sum\limits_{k=1}^{|\mathcal{I}_{i}|}\tilde{p}_{i_{k},i_{j}} \parallel\mathbf{x}_{i_{k},i_{j}}\parallel^{2}}{\leq B_{i_{j}}, ~~\sum\limits_{i_{k} \in \mathcal{K}_{i_{j}}}d_{i_{j},i_{k}}\hat{R}_{i_{k}}||\mathbf{v}_{i_{k},i_{j}}||^{2}}{\leq P_{i_{j}},~~ \mathbf{v}_{i_{k},i_{j}}}{=\mathbf{x}_{i_{k},i_{j}}}
      \addConstraint{\sqrt{w_{i_{k}}}\mathbf{u}_{i_{k}}^{H}\mathbf{H}_{i_{k},i_{j}}\mathbf{v}_{i_{m},i_{j}}}{= X_{i_{k}}^{i_{m},i_{j}}, ~~ \sqrt{w_{i_{k}}}\mathbf{u}_{i_{k}}^{H}\mathbf{H}_{i_{k},q_{m}}\mathbf{v}_{q_{\ell},q_{m}}}{= X_{i_{k}}^{q_{\ell},q_{m}}}
    \end{mini}
    \end{figure*}
\normalsize
~Then, we form the augmented Lagrangian $\mathcal{L}_{\rho}(\mathbf{v},\mathbf{x},\mathbf{X};\boldsymbol{\lambda}_{i_{k}},\mathbf{z}_{i_{k}})$ as follows

\footnotesize
\begin{align}\label{AugmentedLg}
\nonumber
&\mathcal{L}_{\rho}(\mathbf{v},\mathbf{x},\mathbf{X};\boldsymbol{\lambda}_{i_{k}},\mathbf{z}_{i_{k}}) = \\ \nonumber &\sum\limits_{i = 1}^{M} \sum\limits_{k =1}^{|\mathcal{I}_{i}|}\sum\limits_{n=1}^{F}\bar{p}_{i_{k},f_{n}}\big(|\sqrt{ w_{i_{k}}} - \sum\limits_{j = 1}^{|\mathcal{Q}_{i}|}\bar{p}_{i_{k},f_{n}}^{2}\mathbf{C}_{i_{k}}(f_{n},i_{j})\tilde{p}_{i_{k},i_{j}}\beta_{i_{k},i_{j}} X_{i_{k}}^{i_{k},i_{j}}|^{2} \\ \nonumber &+ \sum\limits_{n=1}^{F}\sum\limits_{(\ell, q)\neq (k,i)}\sum\limits_{j =1}^{|\mathcal{Q}_{q}|}\bar{p}_{q_{\ell},f_{n}}^{2}\mathbf{C}_{q_{\ell}}(f_{n},q_{j})\tilde{p}_{q_{\ell},q_{j}}\beta_{i_{k},q_{j}}|X_{i_{k}}^{q_{\ell},q_{j}}|^{2}\big)\\ \nonumber
&+ \text{Re}\big(\sum\limits_{i = 1}^{M} \sum\limits_{k =1}^{|\mathcal{I}_{i}|}\sum\limits_{\substack{m=1 \\ m\neq k}}^{|\mathcal{I}_{i}|}\sum\limits_{j = 1}^{|\mathcal{Q}_{i}|}\langle  \sqrt{w_{i_{k}}} \mathbf{u}_{i_{k}}^{H}\mathbf{H}_{i_{k},i_{j}}\mathbf{v}_{i_{m},i_{j}} - X_{i_{k}}^{i_{m},i_{j}},\lambda_{i_{k}}^{i_{m},i_{j}} \rangle \big)\\ \nonumber
&+\frac{\rho}{2}\sum\limits_{i = 1}^{M} \sum\limits_{k =1}^{|\mathcal{I}_{i}|}\sum\limits_{\substack{m=1 \\ m\neq k}}^{|\mathcal{I}_{i}|}\sum\limits_{j =1}^{|\mathcal{Q}_{i}|}| \sqrt{w_{i_{k}}}\mathbf{u}_{i_{k}}^{H}\mathbf{H}_{i_{k},i_{j}}\mathbf{v}_{i_{m},i_{j}} - X_{i_{k}}^{i_{m},i_{j}}|^{2} \\ \nonumber
&+\text{Re}\big(\sum\limits_{i = 1}^{M} \sum\limits_{k =1}^{|\mathcal{I}_{i}|}\sum\limits_{\substack{q=1 \\ q\neq i}}^{M}\sum\limits_{\ell=1}^{|\mathcal{I}_{q}|}\sum\limits_{m =1}^{|\mathcal{Q}_{q}|}\langle  \sqrt{w_{i_{k}}} \mathbf{u}_{i_{k}}^{H}\mathbf{H}_{i_{k},q_{m}}\mathbf{v}_{q_{\ell},q_{m}} - X_{i_{k}}^{q_{\ell},q_{m}}, \lambda_{i_{k}}^{q_{\ell},q_{m}}\rangle\big) \\ \nonumber
&+ \frac{\rho}{2}\sum\limits_{i = 1}^{M} \sum\limits_{k =1}^{|\mathcal{I}_{i}|}\sum\limits_{\substack{q=1 \\ q\neq i}}^{M}\sum\limits_{\ell=1}^{|\mathcal{I}_{q}|}\sum\limits_{m =1}^{|\mathcal{Q}_{q}|}| \sqrt{w_{i_{k}}}\mathbf{u}_{i_{k}}^{H}\mathbf{H}_{i_{k},q_{m}}\mathbf{v}_{q_{\ell},q_{m}} - X_{i_{k}}^{q_{\ell},q_{m}}|^{2}
\\ \nonumber
&+ \text{Re}\big(\sum\limits_{i =1}^{M}\sum\limits_{k=1}^{|\mathcal{I}_{i}|}\sum\limits_{j=1}^{ |\mathcal{Q}_{i}|}\langle \mathbf{v}_{i_{k},i_{j}} - \mathbf{x}_{i_{k},i_{j}},\mathbf{z}_{i_{k},i_{j}}\rangle\big)\\
&+ \frac{\rho}{2}\sum\limits_{i =1}^{M}\sum\limits_{k=1}^{|\mathcal{I}_{i}|}\sum\limits_{j=1}^{|\mathcal{Q}_{i}|}\|\mathbf{v}_{i_{k},i_{j}} - \mathbf{x}_{i_{k},i_{j}}\|^{2}
\end{align}
\normalsize
where $\rho$ is the penalty parameter, and $\small \boldsymbol{\lambda}_{i_{k}} \triangleq \{\lambda_{i_{k}}^{i_{m},i_{j}}|m,k \in \mathcal{I}_{i},~ j \in \mathcal{Q}_{i}\}$ and $\small \mathbf{z}_{i_{k}} \triangleq \{\mathbf{z}_{i_{k},i_{j}}\in \mathbb{C}^{N_{t}^{i_{j}}}|k \in \mathcal{I}_{i}, ~j \in \mathcal{Q}_{i}\}$ are the scaled dual variables for the last three sets of equality constraints.
\raggedbottom

The ADMM approach consists of three steps. First, minimizing the augmented Lagrangian $\mathcal{L}_{\rho}$ over the decision variables $\mathbf{v}$, while assuming all the other variables are given at their current values. Second,  minimizing the augmented Lagrangian $\mathcal{L}_{\rho}$ over the decision variables $\{\mathbf{x},\mathbf{X}\}$, assuming the rest of variables are given and fixed. While the latter is a constrained optimization problem the former is an unconstrained one. The last step consists of a simple dual update. 
Therefore, assuming $\mathbf{x}$ and $\mathbf{X}$ are given, the optimization problem with respect to $\mathbf{v}$ can be expressed as

\footnotesize
\begin{equation}
\begin{aligned}\label{Eq18}
& \underset{\mathbf{v}}{\text{minimize}}
& & \mathcal{L}_{\rho}(\mathbf{v},\mathbf{x},\mathbf{X};\boldsymbol{\lambda}_{i_{k}},\mathbf{z}_{i_{k}}),
\end{aligned}
\end{equation}
\normalsize
which can be decomposed into $i = 1, \ldots, M,~k = 1, \ldots, |\mathcal{I}_{i}|,~ j = 1, \ldots, |\mathcal{Q}_{i}|$
\footnotesize
\begin{equation}
\begin{aligned}\label{Eq20}
 \underset{\mathbf{v}_{i_{k},i_{j}}}{\text{minimize}}
&~~ f(\mathbf{v}_{i_{k},i_{j}})
\end{aligned}
\end{equation}
\normalsize
where $f(\mathbf{v}_{i_{k},i_{j}})$ is defined as

\footnotesize
\begin{align}\nonumber
&f(\mathbf{v}_{i_{k},i_{j}}) \triangleq \text{Re}\big(\sum\limits_{\substack{m=1 \\ m\neq k}}^{|\mathcal{I}_{i}|}  \lambda_{i_{m}}^{i_{k},i_{j}}\sqrt{w_{i_{m}}} \mathbf{u}_{i_{m}}^{H}\mathbf{H}_{i_{m},i_{j}}\mathbf{v}_{i_{k},i_{j}}\big)+\frac{\rho}{2}\sum\limits_{\substack{m=1 \\ m\neq k}}^{|\mathcal{I}_{i}|} |\sqrt{w_{i_{k}}}\\ \nonumber &
\mathbf{u}_{i_{m}}^{H}\mathbf{H}_{i_{m},i_{j}}\mathbf{v}_{i_{k},i_{j}} - X_{i_{m}}^{i_{k},i_{j}}|^{2}+ \text{Re}\big(\mathbf{z}_{i_{k},i_{j}}^{H}(\mathbf{v}_{i_{k},i_{j}} - \mathbf{x}_{i_{k},i_{j}})\big)\\ \nonumber &
+ \frac{\rho}{2}\|\mathbf{v}_{i_{k},i_{j}} - \mathbf{x}_{i_{k},i_{j}}\|^{2}.
\end{align}
\normalsize

Assuming $\mathbf{v}$ is given, the constrained optimization problem with respect to $\{\mathbf{x},\mathbf{X}\}$ can be expressed as

\small
\begin{align}\label{Eq19}
& \underset{\mathbf{x},\mathbf{X}}{\text{minimize}}
& & \mathcal{L}_{\rho}(\mathbf{v},\mathbf{x},\mathbf{X};\boldsymbol{\lambda}_{i_{k}},\mathbf{z}_{i_{k}}),\\ \nonumber
& \text{subject to}
&& \sum\limits_{k=1}^{|\mathcal{I}_{i}|}\tilde{p}_{i_{k},i_{j}} \parallel\mathbf{x}_{i_{k},i_{j}}\parallel^{2}\leq P_{i_{j}},\quad\forall i \in \mathcal{M}, j \in \mathcal{Q}_{i}\\\nonumber
&&&\sum\limits_{i_{k} \in \mathcal{K}_{i_{j}}}d_{i_{j},i_{k}}\hat{R}_{i_{k}}||\mathbf{v}_{i_{k},i_{j}}||^{2}\leq P_{i_{j}},\quad,\forall i \in \mathcal{M}, j \in \mathcal{Q}_{i}
\end{align}
\normalsize
which can be decomposed into 

\small
\begin{equation}
\begin{aligned}\label{Eq21}
 \underset{X_{i_{k}}^{i_{k},i_{j}}}{\text{minimize}}&
~~ g(X_{i_{k}}^{i_{k},i_{j}})
\end{aligned}
\end{equation}
\normalsize
where $g(X_{i_{k}}^{i_{k},i_{j}})$ is defined as

\small
\begin{align}\nonumber
  &g(X_{i_{k}}^{i_{k},i_{j}}) \triangleq -\sum\limits_{\substack{m=1 \\ m\neq k}}^{|\mathcal{I}_{i}|}\sum\limits_{j=1}^{|\mathcal{Q}_{i}|} X_{i_{m}}^{i_{k},i_{j}}\lambda_{i_{m}}^{i_{k},i_{j}}\\ \nonumber &+\sum\limits_{n=1}^{F}\bar{p}_{i_{k},f_{n}}\big(|\sqrt{ w_{i_{k}}} - \sum\limits_{j = 1}^{|\mathcal{Q}_{i}|}\bar{p}_{i_{k},f_{n}}^{2}\mathbf{C}_{i_{k}}(f_{n},i_{j})\tilde{p}_{i_{k},i_{j}}\beta_{i_{k},i_{j}} X_{i_{k}}^{i_{k},i_{j}}|^{2} \big)\\ \nonumber &+\frac{\rho}{2}\sum\limits_{\substack{m=1 \\ m\neq k}}^{|\mathcal{I}_{i}|}\sum\limits_{j=1}^{|\mathcal{Q}_{i}|}|\sqrt{w_{i_{m}}}\mathbf{u}_{i_{m}}^{H}\mathbf{H}_{i_{m},i_{j}}\mathbf{v}_{i_{k},i_{j}} - X_{i_{m}}^{i_{k},i_{j}}|^{2},
\end{align}
\normalsize
and 

\footnotesize
\begin{equation}
\begin{aligned}\label{Eq22}
 \underset{\mathbf{x}_{i_{k},i_{j}}}{\text{minimize}}&
~~ h(\mathbf{x}_{i_{k},i_{j}})
\\
\text{subject to}
&~~\sum\limits_{k=1}^{|\mathcal{I}_{i}|}\tilde{p}_{i_{k},i_{j}} \parallel\mathbf{x}_{i_{k},i_{j}}\parallel^{2}\leq P_{i_{j}},
\\&\sum\limits_{i_{k} \in \mathcal{K}_{i_{j}}}d_{i_{j},i_{k}}\hat{R}_{i_{k}}||\mathbf{v}_{i_{k},i_{j}}||^{2}\leq P_{i_{j}},\quad,\forall i \in \mathcal{M}, j \in \mathcal{Q}_{i}
\end{aligned}
\end{equation}
\normalsize
where $h(\mathbf{x}_{i_{k},i_{j}})$ is defined as follows

\small
\begin{align}\nonumber
h(\mathbf{x}_{i_{k},i_{j}})&\triangleq \text{Re}\big(\sum\limits_{k=1}^{|\mathcal{I}_{i}|}\sum\limits_{j=1}^{| \mathcal{Q}_{i}|}\langle \mathbf{v}_{i_{k},i_{j}} - \mathbf{x}_{i_{k},i_{j}},\mathbf{z}_{i_{k},i_{j}}\rangle \big)\\ \nonumber &+  \frac{\rho}{2}\sum\limits_{k=1}^{|\mathcal{I}_{i}|}\sum\limits_{j=1}^{|\mathcal{Q}_{i}|}\|\mathbf{v}_{i_{k},i_{j}} - \mathbf{x}_{i_{k},i_{j}}\|^{2}
\end{align}
\normalsize
Note that all the three subproblems (\ref{Eq20}) and (\ref{Eq21})--(\ref{Eq22}) are convex problems and can be solved efficiently. After calculating each variable at the $m+1$ iteration, we will update the $\boldsymbol{\lambda}_{i_{k}}$ as discussed earlier.

\subsection{\textbf{Long-term Caching Placement:}}
As discussed earlier, the caching problem can be decoupled among all the clusters in the aforementioned MEC network. Specifically, the caching problem for the $i$-th cluster can be expressed as the stochastic optimization problem (\ref{eq_34}) (since the average rate is in the form of expectation over all possible channel realizations and user requests), where $\mu_{i_{j},f_{n}}= \sum_{k = 1}^{|\mathcal{I}_{i}|}\bar{p}_{i_{k},f_{n}}d_{i_{j},i_{k}}$ represents the amount of backhaul savings due to storing $f_{n}$-th file in the $i_{j}$-th BS's cache. Maximization of $\sum_{n = 1}^{F}\mu_{i_{j},f_{n}}\mathbf{C}_{i}(f_{n},i_{j})$ for a simple SISO case in which both BSs and UEs are equipped with a single antenna, and each UE is only connected with one BS is considered in \cite{Dai-2016}. This maximization poses the following question: which files should be cached in the $i_{j}$-th BS to achieve the backhaul reduction $\mu_{i_{j},f_{n}}$? Under the condition that all files have the same size, the optimal solution to maximize $\sum_{n = 1}^{F}\mu_{i_{j},f_{n}}\mathbf{C}_{i}(f_{n},i_{j})$ is caching the $s_{i_{j}}$ files that make the largest backhaul reduction; i.e.,

\begin{figure*}
\footnotesize
\begin{equation}
\begin{alignedat}{2}\label{eq_34}
\mathcal{P}_{3}:~  &\underset{\mathbf{C}_{i}}{\text{maximize}}
     &\bigg(&\lambda\sum\limits_{i = 1}^{M}\sum\limits_{k=1}^{|\mathcal{I}_{i}|} \sum\limits_{n=1}^{F}\bar{p}_{i_{k},f_{n}}R_{i_{k},f_{n}} + (1-\lambda)\sum\limits_{i = 1}^{M}\sum\limits_{n = 1}^{F}\sum\limits_{j=1}^{|\mathcal{Q}_{i}|}\mu_{i_{j},f_{n}}\mathbf{C}_{i}(f_{n},i_{j})\bigg)\\
& \text{subject to}
&&\sum\limits_{n = 1}^{F}\mathbf{C}_{i}(f_{n},i_{j})\ell_{f_{n}}\leq s_{i_{j}}, ~\forall i \in \mathcal{M}, ~ j \in \mathcal{Q}_{i}
\end{alignedat}
\end{equation}
\end{figure*}
\normalsize

\small
\begin{align}
\nonumber
\mathbf{C}_{i}^{\ast}(f_{n},i_{j}) =
\begin{cases}
    1   ,    & \quad \text{if } \mu_{i_{j},f_{n}} \in \{\mu_{i_{j},f_{1}}, \ldots, \mu_{i_{j},f_{s_{i_{j}}}}\}\\
    0   ,    & \quad \text{Otherwise} \\
\end{cases}
\end{align}
\normalsize
where $\mu_{i_{j},s}$ is the $s$-th item which is more requested in the list of $\mu_{i_{j},f_{n}}$ \cite{Dai-2016}. If each content has different sizes, the aforementioned caching strategy at each BS becomes a knapsack problem \cite{Garey-1990} that can be solved using dynamic programming. Here, we consider a more general case than the one discussed in \cite{Dai-2016}, by considering the multi-cluster multi-user MIMO network in which each UE can be associated with a cluster of BSs.  Consequently, the optimization problem becomes more complicated and the objective function can be written as (\ref{Eqn_O})
\begin{figure*}[b]
\footnotesize
\begin{align}\label{Eqn_O}
O(\mathbf{C}_{i},\mathbf{C}_{-i}) &\triangleq\sum\limits_{i = 1}^{M}O_{i}(\mathbf{C}_{i},\mathbf{C}_{-i})= \sum\limits_{i = 1}^{M}
\bigg(\lambda\sum\limits_{k = 1}^{|\mathcal{I}_{i}|}\sum\limits_{n = 1}^{F}\bar{p}_{i_{k},f_{n}} \mathbb{E}\bigg[\log_{2}(\det \mathbf{I} + \mathbf{M}_{i}(\mathbf{C}_{i})\times \mathbf{N}_{i}^{-1}(\mathbf{C}_{-i}))\bigg]+(1-\lambda)\sum\limits_{n = 1}^{F}\sum\limits_{j=1}^{|\mathcal{Q}_{i}|}\mu_{i_{j},f_{n}}\mathbf{C}_{i}(f_{n},i_{j})\bigg)
\end{align}
\end{figure*}
\normalsize
where $\mathbf{C}_{-i} \triangleq (\mathbf{C}_{q})_{q \neq i}$ and $\mathbf{M}_{i}(\mathbf{C}_{i}) = \sum_{j}\bar{p}_{i_{k},f_{n}}^{2}\mathbf{C}_{i}(f_{n},i_{j})d_{i_{j},i_{k}}\tilde{p}_{i_{k},i_{j}}\beta_{i_{k},i_{j}}\mathbf{H}_{i_{k},i_{j}}\mathbf{v}_{i_{k},i_{j}}\mathbf{v}_{i_{k},i_{j}}^{H}\mathbf{H}_{i_{k},i_{j}}^{H}$, and $\mathbf{N}_{i}(\mathbf{C}_{-i}) = \sum_{(\ell, q)\neq (k,i)}\sum_{j}\bar{p}_{q_{\ell},f_{n}}^{2}\mathbf{C}_{q}(f_{n},q_{j})d_{q_{j},q_{\ell}}\tilde{p}_{q_{\ell},q_{j}}$ $\beta_{i_{k},q_{j}}\mathbf{H}_{i_{k},q_{j}}\mathbf{v}_{q_{\ell},q_{j}}\mathbf{v}_{q_{\ell},q_{j}}^{H}\mathbf{H}_{i_{k},q_{j}}^{H}+\sigma^{2}\mathbf{I}.$
%


Since the entries of $\mathbf{C}_{i}$ take binary values $0$ and $1$, the optimization problem falls into a MINLP category, which is usually NP-hard in general \cite{Kleinberg-2005} and non-tractable in practice. Therefore, we are interested in obtaining a near-optimal solution. Inspired by the method proposed in \cite{Nemhauser-1994}, we allow the binary variables to take real values in $[0,1]$, and hence the original MINLP can be relaxed to a non-linear programming problem. Our aim is to introduce a distributed solution method that efficiently computes the stationary solutions of this problem. To do so, we develop a stochastic parallel successive convex approximation (SCA)-based method \cite{SSCA_2016,SSCA_2019} and substitute a series of strongly convex problems for the optimization problem (\ref{eq_34}). The main idea here is approximating the non-convex objective function $O(\mathbf{C}_{i},\mathbf{C}_{-i})$ by a suitable convex approximation. To be specific, the aim of BSs in each cluster is to choose a feasible cache placement matrix $\mathbf{C}_{i}$ that maximizes the objective function $O(\mathbf{C}_{i},\mathbf{C}_{-i})$ assuming the strategy profile $\mathbf{C}_{-i}$ is given. Inspired by the scheme introduced in \cite{Scutari-2014,Mosleh-2016}, our method is based on solving a sequence of parallel convex problems, i.e., one for each cluster. Each of these convex problems is obtained by maintaining the convex structure of the utility function while linearizing the rest around $\bar{\mathbf{C}_{i}}$. To isolate the inter-cluster and intra-cluster interferences that make $O(\mathbf{C}_{i},\mathbf{C}_{-i})$ nonconvex, we define the utility function of the clusters other than the $i$-th cluster as

\footnotesize
\begin{align}\nonumber
f_{i}(\mathbf{C}_{i},\mathbf{C}_{-i}) &= \sum_{s \neq i}\bigg(\lambda\sum_{k = 1}^{|\mathcal{I}_{s}|}\sum_{n = 1}^{F}\bar{p}_{s_{k},f_{n}} \mathbb{E}\bigg[\log_{2}\det( \mathbf{I} + \mathbf{M}_{s}(\mathbf{C}_{s})\\ \nonumber &\mathbf{N}_{s}^{-1}(\mathbf{C}_{-s}))\bigg]+(1-\lambda)\sum\limits_{n = 1}^{F}\sum\limits_{j=1}^{|\mathcal{Q}_{s}|}\mu_{s_{j},f_{n}}\mathbf{C}_{s}(f_{n},s_{j})\bigg).
\end{align}
\normalsize

Making the objective function convex can be done by keeping the convex part $O_{i}(\mathbf{C}_{i},\mathbf{C}_{-i})$ while linearizing the nonconvex part $f_{i}(\mathbf{C}_{i},\mathbf{C}_{-i})$. Therefore, we use the first order Taylor series expansion of the function $f_{i}(\mathbf{C}_{i},\mathbf{C}_{-i})$. 
that is given by:
\begin{align}\nonumber
f_{i}(\mathbf{C}_{i},\mathbf{C}_{-i})\approx f_{i}(\bar{\mathbf{C}}_{i},\mathbf{C}_{-i}) + f_{i}^{'}(\mathbf{C}_{i},\mathbf{C}_{-i})|_{\mathbf{C}_{i} = \bar{\mathbf{C}}_{i}}(\mathbf{C}_{i} - \bar{\mathbf{C}}_{i}),
\end{align}
where $\mathbf{C}_{i} -\bar{\mathbf{C}}_{i}$ is small, thus the higher order terms can be neglected so that the Taylor’s expansion truncated to the first order. Recalling that $d/dx(\log_{a}(f(x))) = f(x)^{'}/(f(x)\ln a)$, the first-order differential is given by
\footnotesize
\begin{align}
\nonumber
&f_{i}^{'}(\mathbf{C}_{i},\mathbf{C}_{-i}) = \sum\limits_{s \neq i}\bigg(\lambda\sum\limits_{k = 1}^{|\mathcal{I}_{s}|}\sum\limits_{n = 1}^{F} \frac{\xi\bar{p}_{s_{k},f_{n}}}{h(\mathbf{C}_{i})} \times \\ \nonumber & \frac{-\sum\limits_{j =1}^{|\mathcal{Q}_{s}|}\bar{p}_{s_{k},f_{n}}^{2}\mathbf{C}_{s}(f_{n},s_{j})d_{s_{j},s_{k}}\tilde{p}_{s_{k},s_{j}}\beta_{s_{k},s_{j}}\mathbb{E}\{|\mathbf{H}_{s_{k},s_{j}}\mathbf{v}_{s_{k},s_{j}}|^{2}\}}{\big(\sum\limits_{j =1}^{|\mathcal{Q}_{q}|}\bar{p}_{q_{\ell},f_{n}}^{2} \mathbf{C}_{q}(f_{n},q_{j})d_{q_{j},q_{\ell}}\tilde{p}_{q_{\ell},q_{j}}\beta_{s_{k},q_{j}}\mathbb{E}\{|\mathbf{H}_{s_{k},q_{j}}\mathbf{v}_{q_{\ell},q_{j}}|^{2}\}+ \sigma^{2}\mathbf{I}\big)^{2}}\bigg)
\end{align}
\normalsize
 where $\xi = \sum\limits_{j =1}^{|\mathcal{Q}_{i}|}\bar{p}_{i_{\ell},f_{n}}^{2}   d_{i_{j},i_{\ell}}\tilde{p}_{i_{\ell},i_{j}}\beta_{s_{k},i_{j}}\mathbb{E}\{|\mathbf{H}_{s_{k},i_{j}}\mathbf{v}_{iq_{\ell},i_{j}}|^{2}\}$ and $h(\mathbf{C}_{i})$ is given by (\ref{eqn_h}).
 \begin{figure*}
 \small
\begin{align}\label{eqn_h}
&h(\mathbf{C}_{i}) \triangleq 1+\bigg(\sum\limits_{j =1}^{|\mathcal{Q}_{s}|}\bar{p}_{s_{k},f_{n}}^{2}\mathbf{C}_{s}(f_{n},s_{j})d_{s_{j},s_{k}}\tilde{p}_{s_{k},s_{j}}\beta_{s_{k},s_{j}}\mathbb{E}\{|\mathbf{H}_{s_{k},s_{j}}\mathbf{v}_{s_{k},s_{j}}|^{2}\}\bigg) \times\\\nonumber &\bigg(\sum\limits_{(\ell,q)\neq (k,s)}\sum\limits_{j =1}^{|\mathcal{Q}_{q}|}\bar{p}_{q_{\ell},f_{n}}^{2}  \mathbf{C}_{q}(f_{n},q_{j})d_{q_{j},q_{\ell}}\tilde{p}_{q_{\ell},q_{j}}\beta_{s_{k},q_{j}}\mathbb{E}\{|\mathbf{H}_{s_{k},q_{j}}\mathbf{v}_{q_{\ell},q_{j}}|^{2}\}+ \sigma^{2}\mathbf{I}\bigg)^{-1}
\end{align}
\end{figure*}
\normalsize
By keeping only the linear term in the Taylor’s expansion of $f_{i}(\mathbf{C}_{i},\mathbf{C}_{-i})$ around $\bar{\mathbf{C}}_{i}$ and adding a proximal like regularization term, the objective function in (\ref{eq_34}) can be approximated as

\small
\begin{align}\nonumber
\tilde{O}(\mathbf{C}_{i},\mathbf{C}_{-i}) &= O_{i}(\mathbf{C}_{i},\mathbf{C}_{-i})+ \mathbf{C}_{i}(f_{n},i_{j})f_{i}^{'}(\mathbf{C}_{i},\mathbf{C}_{-i})|_{\mathbf{C}_{i} = \bar{\mathbf{C}}_{i}} \\ \nonumber &-\frac{\tau_{i}}{2}|\mathbf{C}_{i}(f_{n},i_{j}) - \bar{\mathbf{C}}_{i}(f_{n},i_{j})|^{2}
\end{align}\normalsize
where $\tau_{i}$ is a given nonnegative constant. We can approximate (\ref{eq_34}) by a set of $|\mathcal{Q}_{i}|$ per cluster problems given by
\begin{equation}\nonumber
\begin{aligned}
& \underset{\mathbf{C}_{i}\in \mathcal{K}}{\text{maximize}}
& & \tilde{O}(\mathbf{C}_{i},\mathbf{C}_{-i}) \\
\end{aligned}
\end{equation}
where $\mathcal{K} \triangleq \{\mathbf{C}_{i}(f_{n}, i_{j})|\sum_{n = 1}^{F}\mathbf{C}_{i}(f_{n},i_{j})\ell_{f_{n}}\leq s_{i_{j}}\}$. Using the proposed algorithm in \cite{Scutari-2014}, for each BS we have the following best response mapping which consists of solving iteratively the sequence of a (strongly) convex optimization problem
  \begin{equation}
\begin{aligned}\label{eqn_35}
\mathbf{C}_{i}^{\ast}(f_{n},i_{j}) = ~
& \underset{\mathbf{C}_{i} \in \mathcal{K}}{\arg \max}
& & \tilde{O}(\mathbf{C}_{i},\mathbf{C}_{-i}) \\
\end{aligned}
\end{equation}
Unlike (\ref{eq_34}), (\ref{eqn_35}) is strongly convex and can be efficiently solved by numerical iterative
algorithms.

\begin{remark}(A Summary of Overall Operation)
Utilizing the timescale separation of the optimization variables, we divide the original solution into short-term content delivery and long-term content placement. While the short-term process consists of user-association and beamforming optimization, the long-term process is composed of cache content placement. MEC incorporates cloud computing capabilities into wireless networks. The network, therefore, benefits from the cloud capability to perform substantial and large-scale resource allocation and interference management. The content placement and the content delivery optimizations are performed in the cloud in order to overcome with any memory requirements. The content placement and the content delivery optimizations are performed in the cloud. The caching placement strategy is adaptive to the channel statistics. As soon as the user request profile changes, the cloud computes the updated cache placement and passes it to the BSs. Then the BSs update their cache. In each channel coherence time, the CSI is acquired from the users through feedback, and the user-association and beamforming vectors are determined based on the instantaneous channel realization.
\end{remark}

\begin{remark} (Computational Complexity)
Since we employ the WMMSE algorithm \cite{Shi-2011} for designing the beamformers, the computational complexity of the beamforming optimization problem is much like the WMMSE, with the difference being that the introduced ADMM-based algorithm decomposes the original large-scale problem into parallel small-scale subproblems. As a result, it needs more complex calculations than the coordinated descent method which is more desirable when the network size is small. However, the computation complexity of the proposed algorithm increases at a slower linear rate with respect to the number of users. The computational complexity of the user-association algorithm is similar to \cite{Shen-2014} and is polynomial in relation to the network size. However, we further lower the complexity of associating a user with a BS by taking content placement into account and excluding the candidate users that increase the backhaul consumption from consideration. The computational complexity of the long-term caching placement is exceedingly low due to the fact that for each realization of the user request profile, the only thing the introduced algorithm needs to do is a simple Jacobi/Gauss-Seidel update. 

To be specific, denote $B$, $K$, $N_{t}$, and $N_{r}$ as the total number of BSs, the total number of users in the system, the number of transmit antenna at each BS, and the number of receive antenna at each user, respectively. We calculate per iteration complexity, where an iteration of algorithm means one round of updating all users’ beamforming and user-association matrices. Under these conditions, each iteration of the proposed algorithm involves the computation of the MMSE receiver vector. To determine this vector in the proposed algorithm, we need to first calculate the covariance matrix of the total received signal at each receiver and then compute their sum. Consequently, the per-iteration complexity of the receive beamformer is $\mathcal{O}(K^{2} B^{2} N_{t} N_{r}^{2} + K^{2}B^{2} N_{r} N_{t}^{2} + K^{2} BN_{t}^{3}+ KBN_{r}^3 )$ . In order to obtain the global computational complexity of the ADMM, again, we calculate the per-iteration complexity which means obtained point should satisfies $\epsilon$-optimality condition $\mathcal{L}_{\rho}(\mathbf{\hat{v}},\mathbf{\hat{x}},\mathbf{\hat{X}};\boldsymbol{\lambda}_{i_{k}},\mathbf{z}_{i_{k}}) - \mathcal{L}_{\rho}(\mathbf{v},\mathbf{x},\mathbf{X};\boldsymbol{\lambda}_{i_{k}},\mathbf{z}_{i_{k}}) \leq \epsilon$. Hence, the per-iteration complexity of the (\ref{Eq18}) and (\ref{Eq19}) is $\mathcal{O}(28N_{t}^{4} )$ where lower order terms are ignored.
\end{remark}

\section*{Simulation Evaluations}
\begin{figure}[!b]
\centering
\includegraphics[width=0.9\linewidth]{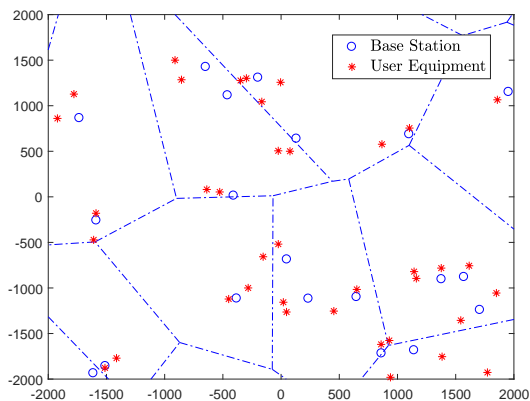}
\caption{A realization of the Cloud-RAN network.}
\label{fig: fig1}
\end{figure}
In this section, we evaluate the performance of the proposed schemes in MEC networks. The setup of our experiments is shown as follows: we simulated a multi-cluster multi-user MEC network in which the locations of the BSs are modeled based on a PPP process with density $\lambda_{B} = 1/(\pi R_{B}^{2}) = 5~\text{BS}/\text{km}^{2}$ which corresponds to an average inter-site distance of $500$ m. Multiple UEs are randomly and uniformly distributed around each BS, excluding an inner circle of $35$ metersas illustrated in Fig. \ref{fig: fig1}. The transmission is subject to interference from all neighboring BSs that do not serve the specific user. The transmit antenna power gain and the transmit power at each BS is set to $10$ dBi and $46$ dBm, respectively. The noise variance at the UE is fixed to $-174$ dBm. System bandwidth is taken as $5$ MHz. We consider a possible antenna configuration in a typical deployment scenario for LTE/LTE-Advanced systems: $4$ transmit and $2$ receive antennas. The simulation is run for $1000$ channel realizations where each channel is uncorrelated Rayleigh fading and each channel element is drawn i.i.d from a complex Gaussian distribution with zero mean and variance $1$. The path-loss is generated using $3$GPP (TR $36.814$) methodology, $\text{PL}(dB) = 148.1 + 37.6\log_{10}(d)$, where $d$ is the distance in kilometers. The log-normal shadowing parameter is assumed to be $8$ dB. The total number of files available in the cloud is considered as $F = 20$. For the sake of simplicity, it is assumed that all files have the same size of one while the caches of the BSs can be filled with the $s = \{1, 2, 4, 8, 10\}$ bits of the most popular files. A Zipf-like distribution with parameter $0.56$ is considered for the file popularity.

\begin{figure}
    \centering
    \includegraphics[width=0.85\linewidth]{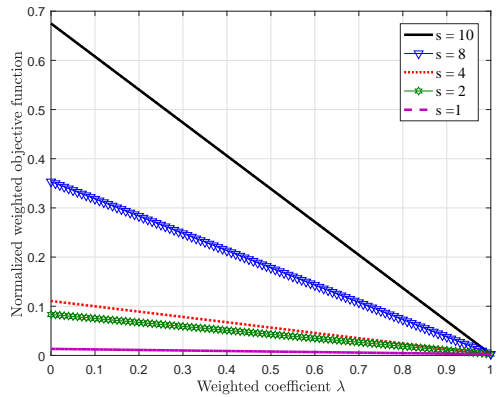}
    \caption{Normalized weighted objective function versus $\lambda$.}
    \label{fig: fig3}
\end{figure}

The influence of the weighted coefficient on the objective introduced function in (\ref{equation20}) is demonstrated in Fig. \ref{fig: fig3}. The results can be interpreted as follows: when $\lambda$ increases the backhaul savings plays a less critical role than the network throughput. Conversely, when $\lambda$ approaches zero the backhaul reduction dominates the objective function. Fig. \ref{fig: fig3} also investigates the influence of the cache size. It shows that the defined weighted objective function of the network with bigger cache size always performs better than the system in which the BSs have a smaller cache. The reason is the smaller the cache sizes, the smaller the portions of the contents can be cached. Therefore, the files must be fetched from the cloud, which increases the backhaul usage. Moreover, when the size of caches grows from $2$ to $8$, an increase of approximately $76.42\%$ in the backhaul savings is observed. Furthermore, $84\%$ additional increase is acquired when the cache sizes enlarge from $1$ to $2$. This shows that even a small size of cache at each BS causes a substantial decrease in the backhaul usage.

\begin{figure}
    \centering
    \includegraphics[width=0.85\linewidth]{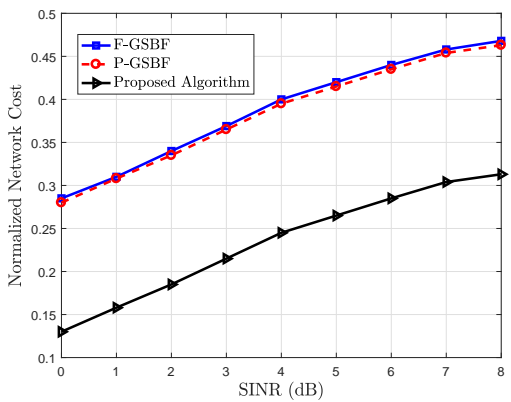}
    \caption{Normalized network cost versus SINR.}
    \label{fig: fig8}
\end{figure}

\begin{figure}
    \centering
    \includegraphics[width=0.85\linewidth]{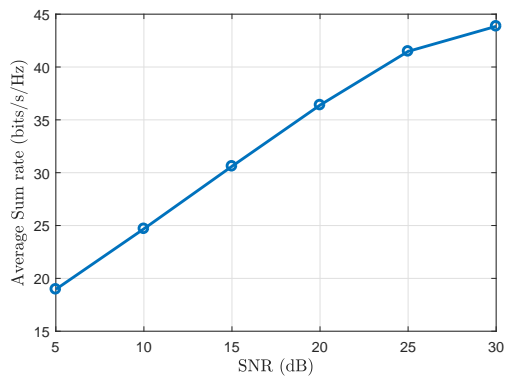}
    \caption{Average sum-rate versus SNR.}
    \label{fig: fig6}
\end{figure}

\begin{figure}
    \centering
    \includegraphics[width=0.85\linewidth]{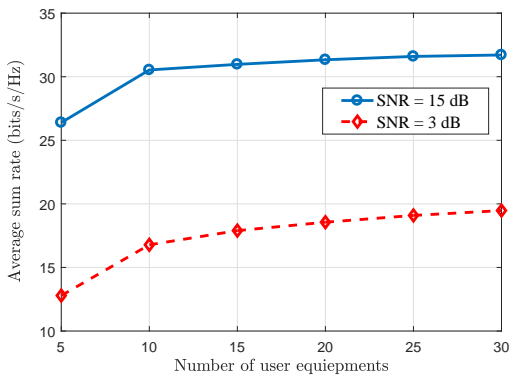}
    \caption{Average sum-rate versus number of UEs.}
    \label{fig: fig5}
\end{figure}

\begin{figure}
    \centering
    \includegraphics[width=0.85\linewidth]{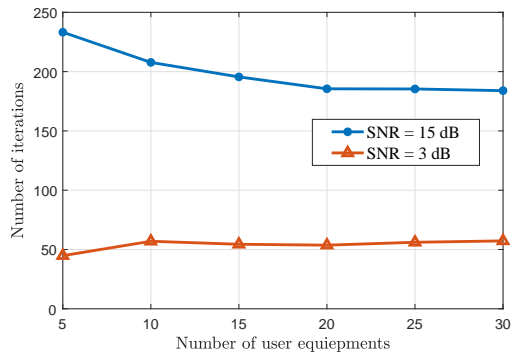}
    \caption{Average number of iterations versus number of UEs.}
    \label{fig: fig7}
\end{figure}

\begin{figure}
    \centering
    \includegraphics[width=0.85\linewidth]{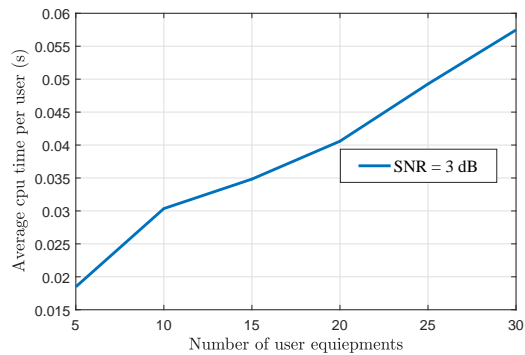}
    \caption{Average CPU time versus number of UEs.}
    \label{fig: fig4}
\end{figure}

In order to compare the performance of the proposed algorithm with some benchmarks, we consider the benchmark algorithm proposed in \cite{Peng-2014} (which is considered as a benchmark by many authors). Note that, despite our interest in joint optimizing of the beamforming, user association, and cache placement, no one to the best of our knowledge has studied maximizing the trade-off between network throughput and backhaul savings in a MIMO multi-user multi-cluster MEC networks. The proposed algorithm in \cite{Peng-2014} considered the joint optimizing beamforming and user association while considering the caching strategy is fixed. Fig. \ref{fig: fig8} demonstrates the performance comparison between our scheme and the ones proposed in \cite{Peng-2014}. A different metric, normalized network cost, is used for the comparison, which is defined as a weighted sum of the backhaul cost and the transmit power cost. As shown in Fig. \ref{fig: fig8}, compared to the full group sparse beamforming algorithm and the partial group sparse beamforming algorithm our introduced algorithm can reduce the network cost, which can interpreted as: taking caching placement into account can be viewed as an additional incentive for backhaul reduction.
The average network throughput versus the SNR and the number of UEs in the MEC network are plotted in Fig. \ref{fig: fig6} and \ref{fig: fig5}, respectively. In Fig. \ref{fig: fig6}, we assume that each BS has the same cache size of $4$ and the network contains $10$ UEs. The algorithm is initialized by choosing a randomly generated feasible point. Moreover, the termination criterion is satisfied when the absolute value of the network throughput error in two consecutive rounds becomes smaller than $10^{-2}$. In Fig. \ref{fig: fig5} we assumed that UEs have the same SNR, due to the fact that co-scheduled UEs usually have similar SNRs in multi-user MIMO operation. The average network throughput is plotted for two SNR values. It is observed that the average sum rate gradually increases when the number of UEs becomes larger.
In addition, the average number of iterations required by
the proposed algorithms to converge versus the number of UEs is plotted in Fig. \ref{fig: fig7}. The average is taken over $1000$ independent channel realizations. It is observed that the algorithm converges in several steps. Moreover, the average CPU time versus the total number of UEs is plotted in Fig. \ref{fig: fig4}. Our experiments were run using MATLAB R$2016$b on a $3.6$ GHz Intel(R) Xeon(R) E$51620$ Processor Cores machine, equipped with $8$ GB of memory. As can be expected, the average CPU time increases with the number of UEs.
\section{Conclusion}
In this paper we introduced a novel iterative algorithm to increase the network throughput and backhaul savings of a multi-cluster multi-user MEC network, by jointly optimizing the user association, caching placement, and beamforming design. The proposed algorithm utilizes the ADMM along with the stochastic SCA-based method and enables all BSs to update their optimization variables in parallel. Simulation results demonstrate that efficiently designing of the content placement along with the content delivery greatly influence the backhaul savings and network throughput.

\bibliographystyle{IEEEtran}
\bibliography{IEEEabrv,MIMO}

\begin{thebibliography}{10}
\providecommand{\url}[1]{#1}
\csname url@samestyle\endcsname
\providecommand{\newblock}{\relax}
\providecommand{\bibinfo}[2]{#2}
\providecommand{\BIBentrySTDinterwordspacing}{\spaceskip=0pt\relax}
\providecommand{\BIBentryALTinterwordstretchfactor}{4}
\providecommand{\BIBentryALTinterwordspacing}{\spaceskip=\fontdimen2\font plus
\BIBentryALTinterwordstretchfactor\fontdimen3\font minus
  \fontdimen4\font\relax}
\providecommand{\BIBforeignlanguage}[2]{{%
\expandafter\ifx\csname l@#1\endcsname\relax
\typeout{** WARNING: IEEEtran.bst: No hyphenation pattern has been}%
\typeout{** loaded for the language `#1'. Using the pattern for}%
\typeout{** the default language instead.}%
\else
\language=\csname l@#1\endcsname
\fi
#2}}
\providecommand{\BIBdecl}{\relax}
\BIBdecl

\bibitem{Cisco_VNI2017}
``Cisco visual networking index: Global mobile data traffic forecast update,
  2017–2022 white paper,'' Cisco Systems, Feb. 2019.

\bibitem{Mosleh-2017}
S.~Mosleh, L.~Liu, H.~Hou, and Y.~Yi, ``Coordinated data assignment: A novel
  scheme for big data over cached cloud-{RAN},'' \emph{IEEE Global
  Communications Conference (GLOBECOM)}, Feb. 2017.

\bibitem{fan2019on}
Q.~{Fan} and N.~{Ansari}, ``On cost aware cloudlet placement for mobile edge
  computing,'' \emph{IEEE/CAA Journal of Automatica Sinica}, vol.~6, no.~4, pp.
  926--937, 2019.

\bibitem{Wang2017a}
S.~{Wang et al.}, ``A survey on mobile edge networks: Convergence of computing,
  caching and communications,'' \emph{IEEE Access}, vol.~5, no. Mar., pp.
  6757--6779, 2017.

\bibitem{etsi2016mobile}
M.~ETSI, ``Mobile edge computing (mec); framework and reference architecture,''
  \emph{ETSI, DGS MEC}, vol.~3, 2016.

\bibitem{Maddah-2013}
M.~A. Maddah-Ali and U.~Niesen, ``Fundamental limits of caching,'' \emph{IEEE
  International Symposium on Information Theory Proceedings (ISIT)}, pp.
  1077--1081, July 2013.

\bibitem{Lau-2013}
A.~Liu and V.~K.~N. Lau, ``Mixed-timescale precoding and cache control in
  cached {MIMO} interference network,'' \emph{{IEEE} Trans. Signal Process.},
  vol.~61, no.~24, pp. 6320--6332, Dec. 2013.

\bibitem{Sengupta-2014}
A.~Sengupta, S.~Amuru, R.~Tandon, R.~M. Buehrer, and T.~C. Clancy, ``Learning
  distributed caching strategies in small cell networks,'' \emph{11th
  International Symposium on Wireless Communications Systems (ISWCS)}, pp.
  1--5, 2014.

\bibitem{Paschos-2016}
G.~Paschos, E.~Bastug, I.~Land, G.~Caire, and M.~Debbah, ``Wireless caching:
  Technical misconceptions and business barriers,''
  \emph{http://arxiv.org/pdf/1602.00173}, Jan. 2016.

\bibitem{Tran-2016}
T.~X. Tran and D.~Pompili, ``Octopus: A cooperative hierarchical caching
  strategy for cloud radio access networks,'' \emph{in Proc. IEEE International
  Conference on Mobile Ad-hoc and Sensor Systems (MASS)}, 2016.

\bibitem{Hajisami-2017}
T.~X. Tran, P.~Pandey, A.~Hajisami, and D.~Pompili, ``Collaborative
  multi-bitrate video caching and processing in mobile-edge computing
  networks,'' \emph{in Proc. IEEE Conference on Wireless On-demand Network
  Systems and Services (WONS)}, 2017.

\bibitem{Mosleh-2016}
S.~Mosleh, L.~Liu, and J.~Zhang, ``Proportional-fair resource allocation for
  coordinated multi-point transmission in {LTE-A}dvanced,'' \emph{{IEEE} Trans.
  Wireless Commun.}, vol.~15, no.~8, pp. 5355--5367, Aug. 2016.

\bibitem{Golrezaei-2014}
N.~Golrezaei, P.~Mansourifard, A.~Molisch, and A.~Dimakis, ``Base station
  assisted device-to-device communications for high-throughput wireless video
  networks,'' \emph{{IEEE} Trans. Wireless Commun.}, vol.~13, no.~7, pp.
  3665--3676, July 2014.

\bibitem{Leung-2014}
X.~Wang, M.~Chen, T.~Taleb, A.~Ksentini, and V.~C.~M. Leung, ``Cache in the
  air: Exploiting content caching and delivery techniques for 5{G} systems,''
  \emph{{IEEE} Commun. Mag.}, vol.~52, no.~2, pp. 131--139, Feb. 2014.

\bibitem{Bastug-2014}
E.~Bastug, M.~Bennis, and M.~Debbah, ``Living on the edge: The role of
  proactive caching in 5{G} wireless networks,'' \emph{{IEEE} Commun. Mag.},
  vol.~52, no.~8, pp. 82--89, Aug. 2014.

\bibitem{Boccardi-2014}
F.~Boccardi, R.~W. Heath, A.~Lozano, T.~L. Marzetta, and P.~Popovski, ``Five
  disruptive technology directions for 5{G},'' \emph{{IEEE} Commun. Mag.},
  vol.~52, no.~2, pp. 74--80, Feb. 2014.

\bibitem{beck2014mobile}
M.~T. Beck, M.~Werner, S.~Feld, and S.~Schimper, ``Mobile edge computing: A
  taxonomy,'' in \emph{Proc. of the Sixth International Conference on Advances
  in Future Internet}.\hskip 1em plus 0.5em minus 0.4em\relax Citeseer, 2014,
  pp. 48--55.

\bibitem{yao2019on}
J.~{Yao}, T.~{Han}, and N.~{Ansari}, ``On mobile edge caching,'' \emph{IEEE
  Communications Surveys Tutorials}, vol.~21, no.~3, pp. 2525--2553, 2019.

\bibitem{Dai-2014}
B.~Dai and W.~Yu, ``Sparse beamforming for limited-backhaul network {MIMO}
  system via reweighted power minimization,'' \emph{in Proc. IEEE Global
  Communications Conference (GLOBECOM)}, Jun. 2014.

\bibitem{Zhuang-2014}
F.~Zhuang and V.~Lau, ``Backhaul limited asymmetric cooperation for {MIMO}
  cellular networks via semidefinite relaxation,'' \emph{{IEEE} Trans. Signal
  Process.}, vol.~62, no.~3, p. 684–693, Feb. 2014.

\bibitem{Peng-2014}
X.~Peng, J.~C. Shen, J.~Zhang, and K.~B. Letaief, ``Joint data assignment and
  beamforming for backhaul limited caching networks,'' \emph{International
  Symposium on Personal, Indoor and Mobile Radio Communication}, pp.
  1370--1374, Sept. 2014.

\bibitem{Yu-2014}
B.~Dai and W.~Yu, ``Sparse beamforming and user-centric clustering for downlink
  cloud radio access network,'' \emph{IEEE Access: Recent Advances in Cloud
  Radio Access Networks}, vol.~2, pp. 1326--1339, Oct. 2014.

\bibitem{huang2020online}
X.~{Huang}, S.~{Zhao}, X.~{Gao}, Z.~{Shao}, H.~{Qian}, and Y.~{Yang}, ``Online
  user-ap association with predictive scheduling in wireless caching
  networks,'' \emph{IEEE Transactions on Mobile Computing},
  DOI:10.1109/TMC.2020.3036876, 2020, early access.

\bibitem{Yu-2016}
H.~Zhou, M.~Tao, E.~Chen, and W.~Yu, ``Content-centric multicast beamforming in
  cache-enabled cloud radio access networks,'' \emph{in Proc. IEEE Global
  Communications Conference (GLOBECOM)}, Feb. 2016.

\bibitem{Tao-2016}
M.~Tao, E.~Chen, H.~Zhou, and W.~Yu, ``Content-centric sparse multicast
  beamforming for cache-enabled cloud {RAN},'' \emph{{IEEE} Trans. Wireless
  Commun.}, vol.~15, no.~9, pp. 6118--6131, Sep. 2016.

\bibitem{Dimakis-2012}
N.~Golrezaei, K.~Shanmugam, A.~G. Dimakis, A.~F. Molisch, and G.~Caire,
  ``Femtocaching: Wireless video content delivery through distributed caching
  helpers,'' \emph{IEEE INFOCOM}, p. 1107–1115, Mar. 2012.

\bibitem{Shanmugam-2013}
K.~Shanmugam, N.~Golrezaei, A.~G. Dimakis, A.~F. Molisch, and G.~Caire,
  ``Femtocaching: Wireless content delivery through distributed caching
  helpers,'' \emph{{IEEE} Trans. Inf. Theory}, vol.~59, no.~12, p. 8402–8413,
  Dec. 2013.

\bibitem{Li-2015}
J.~Li, Y.~Chen, Z.~Lin, W.~Chen, B.~Vucetic, and L.~Hanzo, ``Distributed
  caching for data dissemination in the downlink of heterogeneous networks,''
  \emph{{IEEE} Trans. Commun.}, vol.~63, no.~10, p. 3553–3568, Oct. 2015.

\bibitem{Rezvani2020fairness}
S.~{Rezvani}, N.~{Mokari}, M.~R. {Javan}, and E.~A. {Jorswieck}, ``Fairness and
  transmission-aware caching and delivery policies in ofdma-based hetnets,''
  \emph{IEEE Transactions on Mobile Computing}, vol.~19, no.~2, pp. 331--346,
  2020.

\bibitem{Rong-2013}
Q.~Ye, B.~Rong, Y.~Chen, M.~Al-Shalash, C.~Caramanis, and J.~G. Andrews, ``User
  association for load balancing in heterogeneous cellular networks,''
  \emph{{IEEE} Trans. Wireless Commun.}, vol.~12, no.~6, pp. 2706--2716, Jun.
  2013.

\bibitem{Shen-2014}
K.~Shen and W.~Yu, ``Distributed pricing-based user association for downlink
  heterogeneous cellular networks,'' \emph{{IEEE} J. Sel. Areas Commun.},
  vol.~32, no.~6, pp. 1100--1113, Jun. 2014.

\bibitem{Pantisano-2014}
F.~Pantisano, M.~Bennis, W.~Saad, and M.~Debbah, ``Cache-aware user association
  in backhaul-constrained small cell networks,'' \emph{12th International
  Symposium on Modeling and Optimization in Mobile, Ad Hoc, and Wireless
  Networks (WiOpt)}, pp. 37--42, May 2014.

\bibitem{Poularakis-2014}
K.~Poularakis, G.~Iosifidis, and L.~Tassiulas, ``Approximation algorithms for
  mobile data caching in small cell networks,'' \emph{{IEEE} Trans. Commun.},
  vol.~62, no.~10, p. 3665–3677, Oct. 2014.

\bibitem{Iosifidis-2014}
K.~Poularakis, G.~Iosifidis, A.~Argyriou, and L.~Tassiulas, ``Video delivery
  over heterogeneous cellular networks: Optimizing cost and performance,''
  \emph{in Proc. IEEE Conf. Comput. Commun. (INFOCOM)}, p. 1078–1086, Apr.
  2014.

\bibitem{Naveen-2015}
K.~Naveen, L.~Massoulie, E.~Baccelli, A.~C. Viana, and D.~Towsley, ``On the
  interaction between content caching and request assignment in cellular cache
  networks,'' \emph{in Proc. 5th Workshop Things Cellular, Oper., Appl.
  Challenges, New York, NY, USA}, p. 37–42, 2015.

\bibitem{Dehghan-2015}
M.~Dehghan, A.~Seetharam, B.~Jiang, T.~He, T.~Salonidis, J.~Kurose, D.~Towsley,
  and R.~Sitaraman, ``On the complexity of optimal routing and content caching
  in heterogeneous networks,'' \emph{in Proc. IEEE Conf. Comput. Commun.
  (INFOCOM)}, p. 936–944, Apr. 2015.

\bibitem{Khreishah-2016}
A.~Khreishah, J.~Chakareski, and A.~Gharaibeh, ``Joint caching, routing, and
  channel assignment for collaborative small-cell cellular networks,''
  \emph{{IEEE} J. Sel. Areas Commun.}, vol.~34, no.~8, p. 2275–2284, Aug.
  2016.

\bibitem{Wang-2016}
Y.~Wang, X.~Tao, X.~Zhang, and G.~Mao, ``Joint caching placement and user
  association for minimizing user download delay,'' \emph{IEEE Access: Wireless
  Cachiing Technique for 5G}, vol.~4, pp. 8625--8633, Dec. 2016.

\bibitem{jing2019user}
W.~{Jing}, X.~{Wen}, Z.~{Lu}, and H.~{Zhang}, ``User-centric delay-aware joint
  caching and user association optimization in cache-enabled wireless
  networks,'' \emph{IEEE Access}, vol.~7, pp. 74\,961--74\,972, 2019.

\bibitem{Dai-2016}
B.~Dai and W.~Yu, ``Joint user association and content placement for
  cache-enabled wireless access networks,'' \emph{IEEE International Conference
  on Acoustics, Speech and Signal Processing (ICASSP)}, May 2016.

\bibitem{EDebbah-2013}
E.~Baştuğ, J.~L. Guénégo, and M.~Debbah, ``Proactive small cell networks,''
  \emph{ICT}, pp. 1--5, May 2013.

\bibitem{Ng-2010}
C.~T.~K. Ng and H.~Huang, ``Linear precoding in cooperative {MIMO} cellular
  networks with limited coordination clusters,'' \emph{{IEEE} J. Sel. Areas
  Commun.}, vol.~28, no.~9, pp. 1446--1454, Dec. 2010.

\bibitem{MoslehIA-2016}
S.~Mosleh, J.~D. Ashdown, J.~D. Matyjas, M.~J. Medley, J.~Zhang, and L.~Liu,
  ``Interference alignment for downlink {M}ulti-{C}ell {LTE-A}dvanced systems
  with limited feedback,'' \emph{{IEEE} Trans. Wireless Commun.}, vol.~15,
  no.~12, pp. 8107--8121, Dec. 2016.

\bibitem{Mosleh-2015}
S.~Mosleh, L.~Liu, Y.~Li, and J.~Zhang, ``Interference alignment and
  leakage-based iterative coordinated beam-forming for multi-user {MIMO} in
  {LTE}-advanced,'' \emph{IEEE Globecom Workshops (GC Wkshps)}, Dec. 2015.

\bibitem{Sesia-2011}
S.~Sesia, I.~Toufik, and M.~Baker, ``{LTE}: The {UMTS} long term evolution:
  From theory to practice,'' \emph{2nd ed. Hoboken, NJ, USA: Wiley,}, Aug.
  2011.

\bibitem{Breslau-1999}
L.~Breslau, P.~Cao, L.~Fan, G.~Phillips, and S.~Shenker, ``Web caching and
  zipf-like distributions: Evidence and implications,'' \emph{IEEE Eighteenth
  Annual Joint Conference of the IEEE Computer and Communications Societies
  (INFOCOM’99)}, vol.~1, p. 126–134, 1999.

\bibitem{Hefeeda-2008}
M.~Hefeeda and O.~Saleh, ``Traffic modeling and proportional partial caching
  for peer-to-peer systems,'' \emph{IEEE/ACM Transactions on Networking},
  vol.~16, no.~6, p. 1447–1460, Dec. 2008.

\bibitem{Kleinberg-2005}
J.~Kleinberg and E.~Tardos, ``Algorithm design,'' \emph{Boston, MA:
  Addison-Wesley}, 2012.

\bibitem{Shi-2011}
Q.~Shi, M.~Razaviyayn, Z.~Q. Luo, and C.~He, ``An iteratively weighted {MMSE}
  approach to distributed sum-utility maximization for a {MIMO} interfering
  broadcast channel,'' \emph{{IEEE} Trans. Signal Process.}, vol.~59, no.~9,
  pp. 4331--4340, Sept. 2011.

\bibitem{Ma-2016}
T.~Ma, Q.~Shi, and E.~Song, ``Qos-constrained weighted sum-rate maximization in
  multi-cell multi-user {MIMO} systems: An {ADMM} approach,'' \emph{35th
  Chinese Control Conference (CCC)}, July 2016.

\bibitem{Binbin-2015}
B.~Dai and W.~Yu, ``Backhaul-aware multicell beamforming for downlink cloud
  radio access network,'' \emph{IWCPM Workshop}, Sep. 2015.

\bibitem{Parikh-2011}
S.~Boyd, N.~Parikh, E.~Chu, B.~Peleato, and J.~Eckstein, ``Distributed
  optimization and statistical learning via the alternating direction method of
  multipliers,'' \emph{Foundations and Trends in Machine Learning}, vol.~3,
  no.~1, pp. 1--122, 2011.

\bibitem{Garey-1990}
M.~R. Garey and D.~S. Johnson, \emph{Computers and intractability: A Guide to
  the Theory of $NP$-completeness}, 1990.

\bibitem{Nemhauser-1994}
G.~L. Nemhauser, M.~W.~P. Savelsbergh, and G.~S. Sigismondi, ``{MINTO}, a
  {M}ixed {INT}eger {O}ptimizer,'' \emph{Research Letters, Elsevier}, vol.~15,
  pp. 47--58, 1994.

\bibitem{SSCA_2016}
Y.~Yang, G.~Scutari, D.~P. Palomar, and M.~Pesavento, ``A parallel
  decomposition method for nonconvex stochastic multi-agent optimization
  problems,'' \emph{{IEEE} Trans. Signal Process.}, vol.~64, no.~11, pp.
  2949--2964, Jun. 2016.

\bibitem{SSCA_2019}
A.~Liu, V.~Lau, and B.~Kananian, ``Stochastic successive convex approximation
  for non-convex constrained stochastic optimization,''
  \emph{https://arxiv.org/pdf/1801.08266}, Feb. 2019.

\bibitem{Scutari-2014}
G.~Scutari, F.~Facchinei, P.~Song, D.~P. Palomar, and J.-S. Pang,
  ``Decomposition by partial linearization: Parallel optimization of
  multi-agent systems,'' \emph{{IEEE} Trans. Signal Process.}, vol.~62, no.~3,
  p. 641–656, Feb. 2014.

\end{thebibliography}

\end{document}